\documentclass{article}

\usepackage{arxiv}
\usepackage[utf8]{inputenc} 
\usepackage[T1]{fontenc}    
\usepackage{hyperref}       
\usepackage{url}            
\usepackage{booktabs}       
\usepackage{amsfonts}       
\usepackage{nicefrac}       
\usepackage{microtype}      
\usepackage{lipsum}		
\usepackage{graphicx}
\usepackage{natbib}
\usepackage{doi}
\usepackage{makecell}
\usepackage{multirow}
\usepackage{xcolor}
\usepackage{amsmath}
\usepackage{amssymb}

\begin{document}

\title{REFUGE2 Challenge: A Treasure Trove for Multi-Dimension Analysis and Evaluation in Glaucoma Screening}

\date{} 					

\author{Huihui Fang, Fei Li, Junde Wu, Huazhu Fu, Xu Sun, Jaemin Son, Shuang Yu, Menglu Zhang, \\ Chenglang Yuan, Cheng Bian, Baiying Lei, Benjian Zhao, Xinxing Xu, Shaohua Li, Francisco Fumero, José Sigut, \\ Haidar Almubarak, Yakoub Bazi, Yuanhao Guo, Yating Zhou, Ujjwal Baid, Shubham Innani, Tianjiao Guo, Jie Yang, \\ José Ignacio Orlando, Hrvoje Bogunović, Xiulan Zhang, Yanwu Xu, iChallenge-REFUGE study group\footnotemark[1]}

\renewcommand{\thefootnote}{\fnsymbol{footnote}}
\footnotetext[1]{H.~Fang and F.~Li contributed equally to this work. 

X.~Zhang and Y.~Xu are the corresponding authors (E-mail: zhangxl2@mail.sysu.edu.cn; ywxu@ieee.org). 

H.~Fang, F.~Li, H.~Fu, X.~Sun, J.I.~Orlando, H.~Bogunović, X.~Zhang, and Y.~Xu co-organized the REFUGE2 challenge. 

F.~Li and X.~Zhang are with State Key Laboratory of Ophthalmology, Zhongshan Ophthalmic Center, Sun Yat-sen University, Guangdong Provincial Key Laboratory of Ophthalmology and Visual Science,Guangzhou, China.

H.~Fang, J.~Wu, X.~Sun, and Y.~Xu are with Intelligent Healthcare Unit, Baidu Inc., Beijing, China.

H.~Fu, X.~Xu and S.Li are with the Institute of High Performance Computing, Agency for Science, Technology and Research, Singapore.

J.I.~Orlando is with Yatiris Group, PLADEMA Institute, CONICET, UNICEN, Tandil, Argentina.

H.~Bogunović is with Christian Doppler Lab for Artificial Intelligence in Retina, Department of Ophthalmology and Optometry, Medical University of Vienna, Vienna, Austria.

J.~Son is with VUNO Inc. Seoul, Republic of Korea.

S.~Yu is with Tencent HealthCare, Tencent, Shenzhen, China.

M.~Zhang is with Computer Vision Institute, College of Computer Science and Software Engineering of Shenzhen University, Shenzhen, China.

C.~Yuan is with School of Biomedical Engineering, Health Science Center, Shenzhen University, China.

C.~Bian is with Xiaohe Healthcare, ByteDance, Guangzhou, Guangdong 510000, China.

B.~Lei is with School of Biomedical Engineering, Shenzhen University, China.

B.~Zhao is with College of Computer Science \& Software Engineering, Shenzhen University, China.

F.~Fumero and J.~Sigut are with Department of Computer Science and Systems Engineering, Universidad de La Laguna, Spain.

H.~Almubarak is with Saudi Electronic University, Saudi Arabia.

Y.~Bazi is with King Saud University, Saudi Arabia.

Y.~Guo and Y.~Zhou are with Institute of Automation, Chinese Academy of Sciences, Beijing, China, University of Chinese Academy of Sciences, Beijing, China

U.~Baid and S.~ Innani are with SGGS Institute of Engineering and Technology, India.

T.~Guo is with Institute of Medical Robotics, Shanghai Jiao Tong University, China.

J.~Yang is with Institute of Image Processing and Pattern Recognition, Shanghai Jiao Tong University, China.}



\renewcommand{\shorttitle}{\textit{arXiv} Template}



\maketitle

\begin{abstract}
With the rapid development of artificial intelligence (AI) in medical image processing, deep learning in color fundus photography (CFP) analysis is also evolving. Although there are some open-source, labeled datasets of CFPs in the ophthalmology community, large-scale datasets for screening only have labels of disease categories, and datasets with annotations of fundus structures are usually small in size. In addition, labeling standards are not uniform across datasets, and there is no clear information on the acquisition device. Here we release a multi-annotation, multi-quality, and multi-device color fundus image dataset for glaucoma analysis on an original challenge -- Retinal Fundus Glaucoma Challenge 2nd Edition (REFUGE2). The REFUGE2 dataset contains 2000 color fundus images with annotations of glaucoma classification, optic disc/cup segmentation, as well as fovea localization. Meanwhile, the REFUGE2 challenge sets three sub-tasks of automatic glaucoma diagnosis and fundus structure analysis and provides an online evaluation framework. Based on the characteristics of multi-device and multi-quality data, some methods with strong generalizations are provided in the challenge to make the predictions more robust. This shows that REFUGE2 brings attention to the characteristics of real-world multi-domain data, bridging the gap between scientific research and clinical application.
\end{abstract}


\section{Introduction}
\label{sec:introduction}

Glaucoma is a neurodegenerative disorder characterized by gradual damage to the optic nerve and retinal nerve fiber layers, which results in visual field deficits. Early diagnosis and treatment are critical to prevent irreversible vision loss and ultimately blindness. However, early symptoms of glaucoma are relatively difficult to recognize and detect. Currently, confirming a glaucoma diagnosis involves multiple clinical examinations such as measuring the intraocular pressure using a tonometer, inspecting the integrity of the optic nerve head using an ophthalmoscope or optical coherence tomography, 
and measuring the visual field of the patient. Massive screening for glaucoma in risk populations is therefore prohibitive due to the high cost and complexity of getting an accurate diagnosis. Color fundus photography (CFP), on the other hand, is an easy-to-acquire retinal imaging method that offers a cost-effective opportunity for screening glaucoma (\cite{cen2021automatic}), as it allows visualizing relevant retinal structures such as the fovea, the optic disc (OD) and the optic cup (OC) (\cite{han2021application}). Clinicians frequently use the vertical cup-to-disc ratio (vCDR) as an indicator of the disease, with vCDR values larger than 0.7 being associated with a higher risk of glaucoma (\cite{aung2016asia}). Other signs related to the disease can also be observed in this modality, such as abnormal narrowing of the OD rim, OD hemorrhages, and severe retinal nerve fiber layer defects.

Deep learning models have recently shown to be promising tools to enhance the capabilities of CFP for ocular disease assessment (\cite{yan2020deep,holmberg2020self}). Automated glaucoma detection from CFPs has been actively investigated using convolutional neural networks (CNN) (\cite{li2018efficacy,bajwa2019two,jiang2019jointrcnn}). 
Due to the limited availability of public annotated datasets for glaucoma assessment, some alternative training schemes were introduced to better exploit smaller datasets. For instance, Hemelings et al. (\cite{hemelings2020accurate}) proposed to train a deep ResNet-50 model using active learning (\cite{felder2009active}), which first automatically retrieves useful images from an unlabelled set that are then manually annotated and used to iteratively improve the learned model. Another study trend is focused on segmenting and detecting regions of interest. Various U-Net (\cite{ronneberger2015u}) variants, for instance, have been introduced to accurately segment the OD and OC (\cite{fu2018joint,yu2019robust,wang2019patch}), yielding efficient alternatives to automatically quantify the vCDR. Fovea localization has also been approached using CNN
 (\cite{hasan2021drnet}), although not as extensively as OD/OC segmentation due to the limited public datasets with these annotations. Recently, there have also been studies using deep learning methods to simultaneously implement OC/OD segmentation and glaucoma classification to drive the model to learn deep features that are more suitable for glaucoma diagnosis (\cite{wu2022seatrans, wu2020leveraging}).

\begin{figure}[tb]
\centering
\includegraphics[width=0.88\linewidth]{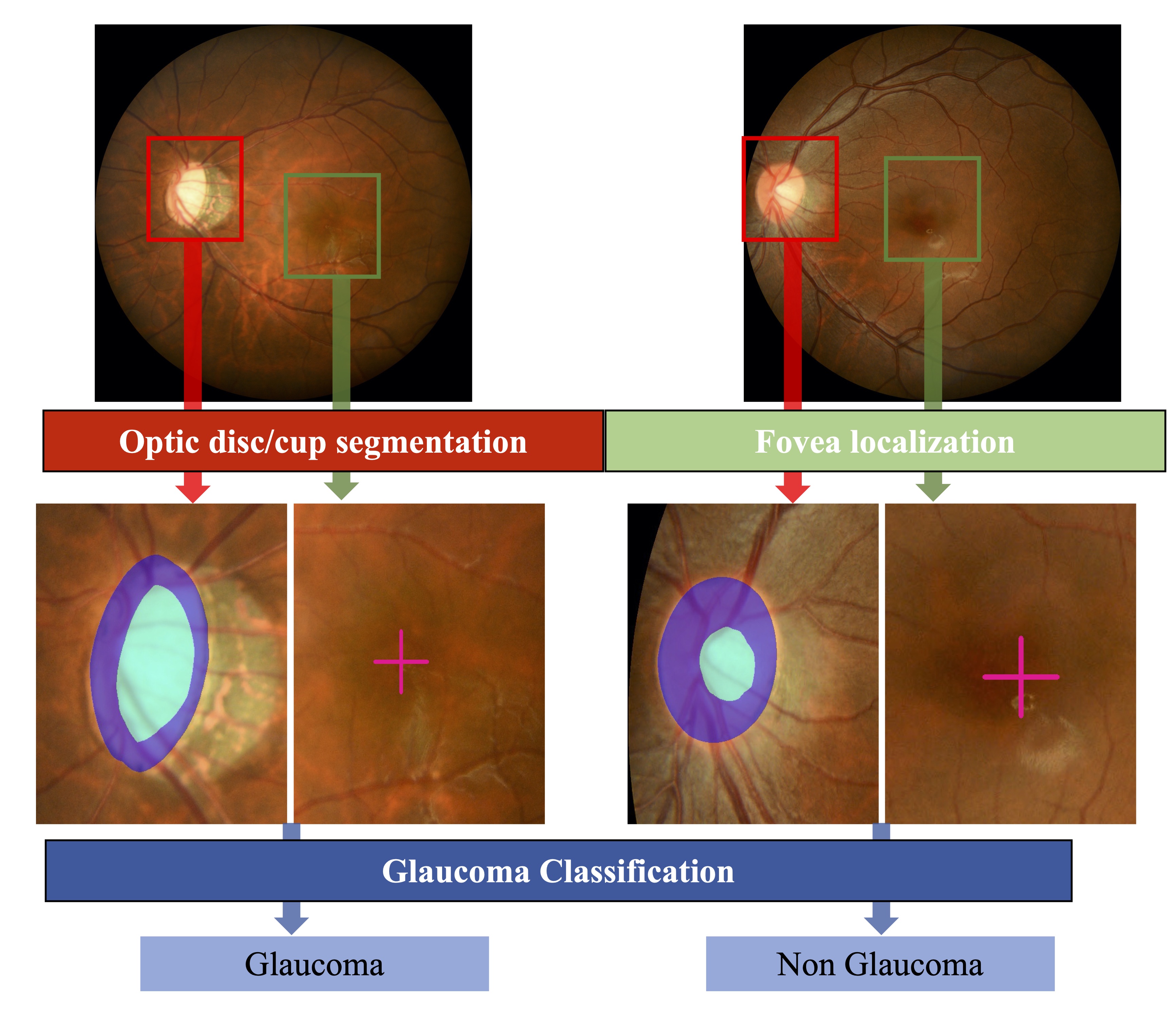}
\caption{REFUGE2 challenge tasks: glaucoma classification, optic disc/cup segmentation, and fovea localization in CFPs.}
\label{fig:subtasks}
\end{figure}

In general, deep neural networks trained using images acquired with a single camera device or from a single population experience a drop in performance when applied to a new dataset for recognizing glaucoma or segmenting the structures of interest. Making the designed models well applicable to the images collected under various situations is of great concern. 
Liu et al. (\cite{liu2019cfea}), for example, proposed a collaborative feature ensembling adaptation method to reduce the effect of the data domain shift. Wang et al. (\cite{wang2019boundary}), on the other hand, used a boundary and entropy-driven adversarial learning alternative to force boundary prediction and mask probability entropy maps obtained on the target domain to be similar to those from the source one, generating more accurate and less uncertain OD/OC segmentations. In Chen’s paper (\cite{chen2021source}), a novel denoised pseudo-labeling approach was introduced to deal with the source-free unsupervised domain adaptation problem. In these studies, the REFUGE1 dataset (\cite{orlando2020refuge}), Drishti-GS (\cite{sivaswamy2014drishti}), RIM-ONE\_r3 (\cite{fumero2011rim}), and their combination were used. 
Although combining multiple datasets for evaluation might offer a good alternative to quantify the robustness of the models to changes in data distribution, each individual set has been annotated following its own standard, which might bias both the training process and the final evaluation outcomes.

\begin{table*}[tbh]
\small
  \centering
  \caption{Summary of public datasets for fundus image analysis, commonly used to train glaucoma assessment algorithms and evaluate their robustness to changes in data distribution.GC-Glaucoma. *The 101,442 images are in the training set of AIROGS dataset, while there are unobtainable 12451 images in the test set with unknown label information, which are unavailable now.
  }
    
    \begin{tabular}{c | c | ccc | ccc}
    \hline
    \multirow{2}[0]{*}{dataset} & \multirow{2}[0]{*}{Device vendors} & \multicolumn{3}{c|}{Num. of images} & \multicolumn{3}{c}{Ground truth labels} \\
    \cline{3-8}
          &       & GC & Non GC & Total & GC classification & Optic disc/cup  & Fovea \\
    \hline
    \makecell{ACRIMA \\(\cite{diaz2019cnns})} & Topcon, IMAGEnet &396 & 309 & 705 & \checkmark & ×/× & × 
    \\\makecell{AIROGS* \\(\cite{de_vente_coen_2021_5793241})} & -&3270 & 98172 & 101,442 & \makecell[c]{\checkmark\\(screening)} & ×/× & × \\
    \makecell{ARIA \\(\cite{zheng2012automated})}   & Zeiss & 0     & 143   & 143   & ×     & \checkmark/×   & \checkmark \\
    \makecell{DIARETDB0 \\(\cite{kauppidiaretdb0})} & -     & -     & -     & 130   & ×     & \checkmark/×   & \checkmark \\
    \makecell{DIARETDB1\\(\cite{kauppi2007diaretdb1}) }& Zeiss & -     & -     & 89    & -     & \checkmark/×   & \checkmark \\
    \makecell{DRIONS-DB \\(\cite{Carmona:2008:ION:1383660.1383874})} & -     & -     & -     & 110   & ×     & \checkmark/×   & × \\
    \makecell{DRISHTI-GS\\(\cite{sivaswamy2014drishti})} & -     & 70    & 31    & 101   & \checkmark     & \checkmark/\checkmark   & × \\
    \makecell{DRIVE\\(\cite{staal2004ridge})} & Canon & -     & -     & 40    & -     & -     & - \\
    \makecell{HEI-MED \\(\cite{giancardo2012hamilton})} & Zeiss & -     & -     & 169   & -     & -     & - \\
    \makecell{HRF\\(\cite{budai2013robust})}  & Canon & 15    & 30    & 45    & \checkmark     & ×/×   & × \\
    \makecell{IDRiD\\(\cite{porwal2018indian})} & KOWA  & 0     & 516   & 516   & ×     & \checkmark/×   & \checkmark \\
    \makecell{MESSIDOR\\(\cite{decenciere2014feedback})}& TOPCON & -     & -     & 1200  & -     & \checkmark/×   & × \\
    \makecell{ORIGA\\(\cite{zhang2010origa})} & Canon & 168   & 482   & 650   & \checkmark     & \checkmark/\checkmark   & × \\
    \makecell{ RIGA\\(\cite{almazroa2018retinal})} & TOPCON, Canon & -     & -     & 750   & ×     & \checkmark/\checkmark   & × \\
    \makecell{RIM-ONE\\(\cite{fumero2011rim})} & Canon & 74    & 85    & 169   & \checkmark     & \checkmark/×   & × \\
    \makecell{SCES\\(\cite{baskaran2015prevalence})} & Canon & 46    & 1630  & 1676  & \makecell[c]{\checkmark\\(screening)} & ×/×   & × \\
    \makecell{STARE\\(\cite{goldbaum2013stare})} & TOPCON & -     & -     & 81    & ×     & \checkmark/×   & \checkmark \\
    
    \makecell{REFUGE1\\(\cite{orlando2020refuge})}& Zeiss, Canon & 120   & 1080  & 1200  & \checkmark     & \checkmark/\checkmark   & \checkmark \\
    \textbf{REFUGE2} & \makecell{\textbf{Canon, Zeiss,}\\\textbf{TOPCON, KOWA}} & \textbf{280} & \textbf{1720} & \textbf{2000} & \textbf{\checkmark} & \textbf{\checkmark/\checkmark} & \textbf{\checkmark} \\
    \hline
    \end{tabular}
  \label{tab:datasets}%
\end{table*}%

Table~\ref{tab:datasets} summarizes all the datasets used for training or testing in the previously described studies. In addition, we supplement the AIROGS dataset (\cite{de_vente_coen_2021_5793241}), the largest dataset available for glaucoma screening. We observe that most of the existing sets have three major deficiencies: (1) limited number of samples, which forces the researchers to combine multiple datasets for their experiments and therefore suffering from mixing multiple labeling standards; (2) no corresponding device information or only single device for the CFPs, which makes it difficult to verify the stability and generalization of an algorithm across images acquired with different cameras; (3) none of them offers simultaneously annotations for all three: glaucoma presence, OD/OC segmentation, and fovea localization.

To overcome these limitations, we release 2,000 CFP dataset acquired with Canon, Zeiss, KOWA, and TOPCON cameras, which includes annotations for OD/OC segmentation, glaucoma classification, and fovea detection. Meanwhile, we host the Retinal Fundus Glaucoma Challenges 2nd Edition (REFUGE2) at the International Conference on Medical Image Computing and Computer Assisted Intervention (MICCAI) 2020 with three sub-tasks (as shown in Fig.~\ref{fig:subtasks}), and release an online evaluation platform to quantify the submitted results due to that the international challenges have become the de facto standard for the comparative evaluation of image analysis algorithms. 

In this article, the main contributions are as follows: 1) We describe in detail the final dataset of 2,000 CFPs. 
To the best of our knowledge, ours is the first public dataset of CFPs acquired with four different camera devices while simultaneously providing annotations about the glaucoma diagnosis, OD/OC segmentation masks and fovea localization. 2) We summarize the high-performing methods of the challenge participating teams, and compare their results on the three proposed sub-tasks, with a special emphasis on their performance of generalization. 3) We discuss the technical implications of using images acquired from multiple devices to achieve domain-agnostic models, the impact of incorporating prior knowledge on them, and the clinical outcomes of these AI methods.

\section{The REFUGE2 challenge}
\label{sect:reguge2}
This section comprises information on challenge organization, challenge sub-tasks, multi-device REFUGE2 dataset, as well as the score rules.

\subsection{Challenge Procedures}
In 2018, to provide the ophthalmic image analysis community with CFPs for glaucoma, we hosted the Retinal Fundus Glaucoma Challenges (REFUGE) at MICCAI, held in Granada, Spain. In REFUGE, we released 1200 CFPs acquired with both Canon and Zeiss cameras, that includes annotations for OD/OC segmentation, glaucoma classification (\cite{orlando2020refuge}). To encourage attention to multi-device data in the clinical practice, we hosted the 2nd edition of REFUGE, called REFUGE2, at virtual edition of MICCAI 2020. In REFUGE2, we further released another 800 densely annotated scans, in this case acquired with two new cameras (KOWA and TOPCON). We also proposed three sub-tasks, namely glaucoma classification, OD/OC segmentation and fovea detection (as shown in Fig.~\ref{fig:subtasks}), and released an online evaluation platform to quantify the submitted results. In terms of the sub-task design, with the exception of glaucoma classification, we continued the OD/OC segmentation task of REFUGE1, because glaucoma has a significant impact on this area. In addition, we added a task of fovea localization because we wanted to explore whether the relationship between fundus structures could help with the single structure analysis. 

REFUGE2 consisted of a preliminary round (online) and a final round (onsite). During the preliminary round, we released the training and online datasets for the model development and evaluation, respectively. The preliminary round ran from July 20 to August 20, during which each team had five opportunities a day to submit their predictions for the online set to the online assessment platform. During the preliminary round, REFUGE2 attracted over 1,300 international participants, with 134 teams submitting over 3,000 prediction results. Finally, 22 teams qualified for the final round. 
The final round was not held onsite due to the pandemic. The onsite set used in the final was sent to each final team over the internet. Teams were given six hours to complete predictions of the onsite dataset and one chance to submit the predictions. All predictions in the final were evaluated offline and the final scores for all teams were calculated according to the scoring rules. The final leaderboards are available at the challenge website. Although the challenge is over now, the data and evaluation framework are still publicly available on \url{ https://refuge.grand-challenge.org/Home2020/}. Future participants are welcome to use our dataset and submit their results on the website and use it for benchmarking their methods.

\subsection{REFUGE2 Dataset}
The REFUGE2 dataset consists of 2,000 retinal CFPs provided by the Zhongshan Ophthalmic Center (Sun Yat-Sen University, China), captured in a darkroom by ophthalmologists and technicians with at least 5 years of collecting experience and stored in JPG format, 8 bits per color channel. CFPs were taken either centered on the OD region, the macular area, or the midpoint between the OD and the macula (with both visible), representing the standard imaging scenario in the clinic. The images in the dataset are CFPs randomly selected from glaucoma and myopia study cohort, and each patient's left and right eyes may be included if the image quality meets the requirements. The personal information of every image was removed for privacy. 1,200 CFPs of the REFUGE2 dataset correspond to the training, online and onsite sets originally released as part of REFUGE1, which were acquired by a Zeiss Visucam 500 camera at a resolution of $2124 \times 2056$  pixels (400 images) and a Canon CR-2 device at a resolution of $1634 \times 1634$ pixels (800 images). As these scans were previously used, 
we included all these scans as a training set for REFUGE2 (see Table~\ref{tab:REFUGE2data}). The remaining 800 images are new and 400 of these scans were acquired using a KOWA device at a resolution of $1940 \times 1940$ pixels, and released as the online set. Another 400 CFPs, which were acquired using a TOPCON TRC-NW400 camera at a resolution of $1848 \times 1848$, were used to build the final onsite set. The dataset is publicly available through the download page of the challenge website and is permitted to be used under the CC BY-NC-ND(Attribution-NonCommercial-NoDerivs) license.

\begin{table*}[tb]
\caption{Summary of the main characteristics of each subset of the REFUGE2 dataset}
\centering
\begin{tabular}{ccccc}
\hline
\multicolumn{1}{c}{\multirow{2}[0]{*}{Characteristics}} & \multicolumn{4}{c}{Subset} \\
\cline{2-5}
      & \multicolumn{2}{c}{Training} & Online & Onsite \\
\hline
Acquisition device    & Zeiss Visucam 500       & Canon CR-2              & KOWA                    & TOPCON TRC-NW400 \\
Resolution            & 2124 $\times$ 2056      & 1634 $\times$ 1634      & 1940 $\times$ 1940      & 1848 $\times$ 1848 \\
Num. images           & \multicolumn{1}{c}{400} & \multicolumn{1}{c}{800} & \multicolumn{1}{c}{400} & \multicolumn{1}{c}{400} \\
Glaucoma/Non glaucoma & 40/360                  & 80/720                  & 80/320                  & 80/320 \\
\hline
\end{tabular}%
\label{tab:REFUGE2data}%
\end{table*}%

\begin{figure}[tb]
\centering
\includegraphics[width=0.83\linewidth]{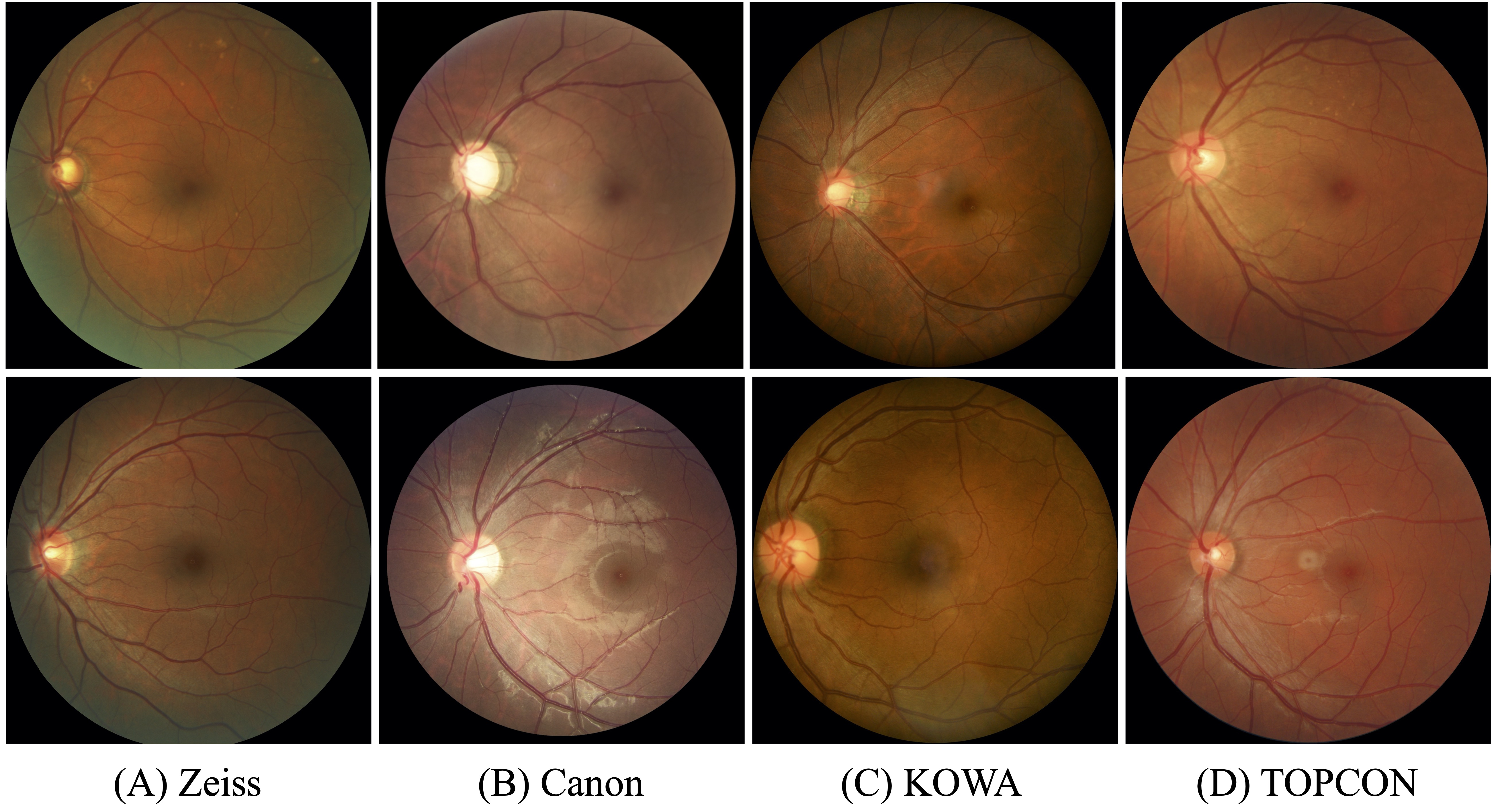}
\caption{Samples collected from the four camera devices. First row: glaucoma, second row: non-glaucoma.}
\label{fig:multi-model}
\end{figure}

\begin{figure}[tb]
\centering
\includegraphics[width=0.7\linewidth]{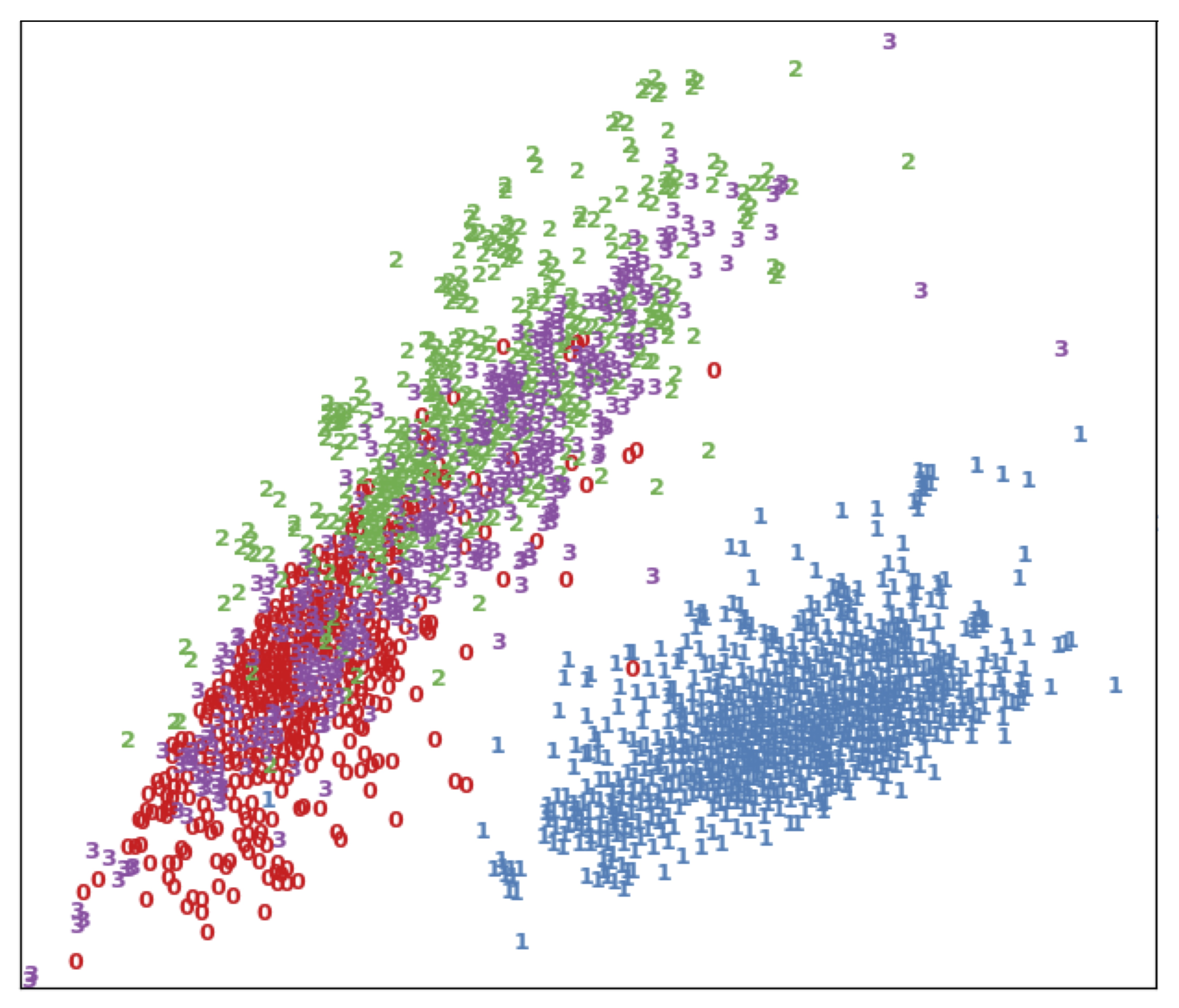}
\caption{t-SNE representations of the original images in the datasets collected from the four camera devices. 0 (red): Zeiss, 1 (blue): Canon, 2 (green): KOWA, 3 (purple): TOPCON.}
\label{fig:data_distribution}
\end{figure}

Fig.~\ref{fig:multi-model} depicts glaucomatous and non-glaucomatous samples acquired using each of the devices. Fig.~\ref{fig:data_distribution} shows a t-distributed Stochastic Neighbor Embedding (t-SNE) (\cite{van2008visualizing}) representation of the 2,000 images, which serves to represent the overall data distribution. The t-SNE was implemented via the scikit-learn package, which is an open-source machine learning toolkit base on Python, and the specific implementation is consistent with that in Fang's work (\cite{fang2022adam}). 
From the figure, we can notice that the 800 Canon training images occupy a different region with respect to the KOWA and TOPCON scans from the online and onsite sets, respectively. Therefore, in REFUGE2, we will select  excellent models with strong generalization ability.

The reference standard for glaucoma classification was obtained from clinical diagnosis results. The manual pixel-wise annotations of the OD/OC segmentation as well as fovea localization were initially delineated by 7 independent ophthalmologists with an average experience of 8 years in the field (ranging from 5 to 10 years), and then fused and checked by a senior specialist with the experience of more than 10 years in the field, which followed the annotating process of REFUGE1 and ADAM challenges (\cite{orlando2020refuge, fang2022adam}).
 
\subsection{Challenge Evaluation}
In the following subsection, we provide the protocol followed to evaluate the challenge results. In task1, the area under the receiver operator characteristics curve (AUC) was calculated as the classification evaluation metric. In task2, Dice coefficient was used to evaluate the segmentation accuracy of OD and OC, meanwhile, the mean absolute error (MAE) was utilized to measure the difference between the vCDR calculated based on the segmented results and those calculated via the reference standard. The vCDR is an important factor in glaucoma assessment, and calculation of vCDR is one of the purposes of the OD/OC segmentation. Hence, we evaluate the effect of OD/OC segmentation by measuring the difference of vCDR. In task3, we utilized the average Euclidean distance (AED) as the evaluation criterion. These three tasks are evaluated in the same way as their counterparts in the REFUGE1 and ADAM challenges (\cite{orlando2020refuge, fang2022adam}).

In tasks 1 and 3, the ranking is based on the AUC and AED metrics. In task 2, each team received three ranks ($R_{disc}, R_{cup}, R_{vCDR}$) from the three evaluation measures based on the mean values over the test set images. The final ranking for the segmentation task was determined by adding the three individual ranks ($R_{total}=R_{disc}+R_{cup}+R_{vCDR}$) with the lower value of $R_{total}$ leading to the higher ranking on the final leaderboard. The comprehensive ranking of the above three tasks was calculated by the following equation:
\begin{equation}
    R = 0.45\times R_{cls}+0.45 \times R_{seg} +0.1 \times R_{loc}
\end{equation}
where $R_{cls}$,$R_{seg}$ and $R_{loc}$ represent the ranks of the aforementioned three tasks. Task 1 and 2 were given higher weights because they are more clinically relevant for glaucoma assessment. This ranking then determined the online or onsite ranking (1=best) of the challenge. In case of a tie, the rank of the classification task has the preference.

Both online and onsite evaluation rankings contribute to the final ranking $R_{final}$:
\begin{equation}
    R_{final} = 0.3 \times R_{online} + 0.7 \times R_{onsite}
\end{equation}
where a higher weight was assigned to the onsite ranking as the online set was involved in the tuning of each team's methods, and the onsite set was a pure blind test set, which can better reflect the generalization ability of the proposed methods, as similarly done in other challenges (\cite{fang2022adam, fuAGEChallengeAngle2020a, orlando2020refuge}). 

\section{Methods}

This section presents the methods designed by the teams which perform well on the REFUGE2 challenge. For task 1, we introduces the VUNO EYE TEAM, MAI, MIG, cheeron, MIAG ULL, ALISR, and EyeStar teams which ranked 1st-4th, 6th, 7th, and 9th in the single task ranking. For task 2, the methods of the cheeron, MAI, VUNO EYE TEAM, EyeStar, MIG, and MIAG ULL teams which ranked 1st-5th, and 8th are introduced. For task 3, we described the methods of the MAI, VUNO EYE TEAM, cheeron, EyeStar, MIG, and ALISR teams which ranked 1st, 3rd, 4th, 6th, 8th, and 9th. The final leaderboards are available on the REFUGE2 website, and the remaining teams ranking in the top 10 of each task gave up participating in this challenge review paper. The general solutions for the challenge and the strategies for processing the multi-device data are presented below. Please refer to the Appendix for specific methods designed by these teams and the other three teams (Pami-G, CBMIBrand, and TeamTiger) that performed better in the semi-final.

\subsection{Challenge General Solutions}

\textit{Classification of clinical glaucoma.} Table~\ref{tab:classication-methods} provides an overview of the methods used by the 7 teams in this task. The VUNO EYE TEAM and EyeStar team proposed the classification methods based on the whole images. The other five teams considered the local information in the OD region, due to the significant variety of the structure and texture in the OD and OC region caused by glaucoma. The MIG team considered both the whole image and the local OD region. The teams preferred EfficientNet (\cite{tan2019efficientnet}), ResNet (\cite{7780459}) and its variants (\cite{gao2019res2net,8100117}), VGGNet (\cite{brusilovsky:simonyan2014very}) and DenseNet (\cite{huang2017densely}) for the classification tasks. A detailed description of the methods designed by these teams is available in the Appendix. 

\begin{table*}[tb]
\renewcommand\arraystretch{3} 
  \centering
  \caption{Methods overview of the 7 participating teams in glaucoma classification task.}
    \resizebox{1.0\textwidth}{!}{%
    \begin{tabular}{ccccc}
    \hline
    \hline
    \textbf{Team}  & \textbf{Input} & \textbf{Architecture} & \textbf{Additional dataset} & \textbf{Highlight} \\
    \hline
    \textbf{\makecell[c]{VUNO EYE \\TEAM}}  &  whole image & EfficientNet & \makecell[c]{Private dataset} & \makecell[c]{Used a pre-trained model which was trained \\by additional samples with 15 lesion labels.} \\
    \textbf{MIG}   & \makecell[c]{ whole image and\\ cropped OD region }& \makecell[c]{Variants \\of ResNet50}  & \makecell[c]{ORIGA, Drishti-GS1,\\ RIM-ONE\_r3, ACRIMA }& \makecell[c]{Ensemble the classification performance of one \\model for processing the whole image and four \\models for processing the OD region image.} \\
    \textbf{MAI}   &  cropped OD region  &  ResNet50  & - & \makecell[c]{A self-supervised task was added to the \\framework and used to update the parameters of \\the feature extraction module during the test phase.} \\
    \textbf{cheeron} & \makecell[c]{cropped OD region} & \makecell[c]{ResNeXt \\ Res2Net} & - & \makecell[c]{The attention mechanism was used in channel and \\ spatial dimensions to increase the network's attention \\to relevant features and suppress unnecessary features.} \\
    \textbf{MIAG ULL} &  cropped OD region  & VGG19 & - &  -\\
    \textbf{ALISR} &  cropped OD region  & CSPResNext50 & - & \makecell[c]{Used a cross-stage partial network, whose complexity \\could be greatly reduced while accuracy was maintained.} \\
    \textbf{EyeStar} & whole image  & DenseNet121 & \makecell[c]{Singapore \\ Epidemiology of \\ Eye Disease} & \makecell[c]{Imposed a new distribution alignment constraint on the \\samples from the SEED domain and the domain of \\the REFUGE2 training set in the shared feature space.} \\
    \hline
    \hline
    \end{tabular}%
  \label{tab:classication-methods}%
  }
\end{table*}%

\textit{Segmentation of optic disc and cup.} A brief summary of the methods adopted by the 6 teams for segmentation task is shown in Table~\ref{tab:methods-segmentation}. The frameworks proposed can be divided in two categories: segmenting OD/OC directly (VUNO EYE TEAM), and segmenting OD/OC from coarse to fine (remaining 5 teams) due to that the OD and OC areas account for a small proportion of the whole fundus image. Most of the base segmentation models used are U-Net (\cite{ronneberger2015u}), while there are also Deeplabv3 (\cite{DBLP:journals/corr/ChenPSA17}) and PSPNet (\cite{8100143}). Feature encoders in the models are mostly replaced by the ResNet (\cite{7780459}) and EfficientNet (\cite{tan2019efficientnet}) structures. Further details can be found in the Appendix.

\begin{table*}[tb]
\renewcommand\arraystretch{3} 
  \centering
  \small
  \caption{Method overview of the 6 participating teams in optic disc and cup segmentation task.}
    \resizebox{1.0\textwidth}{!}{%
    \begin{tabular}{ccccc}
    \hline
    \hline
    \textbf{Team}  & \textbf{Strategy} & \textbf{Architecture} & \textbf{Additional dataset} & \textbf{Highlight} \\
    \hline
    \textbf{cheeron} & \makecell[c]{coarse to fine \\segmentation} & \makecell[c]{U-Net with ResNet as encoder \\for OD coarse segmentation; \\ResUNet for fine segmentation} & \makecell[c]{-} & \makecell[c]{Used deep supervision, atrous\\ spatial pyramidal pooling, and \\test-time augmentation}\\
    \textbf{MAI}   &   \makecell[c]{coarse to fine \\segmentation}   & \makecell[c]{U-Net for OD coarse \\segmentation; DeeplabV3+ \\for precise segmentation} & \makecell[c]{-}     & \makecell[c]{Utilized classical unsupervised\\ domain adaptation strategy} \\
    \textbf{\makecell[c]{VUNO EYE \\TEAM}} & \multicolumn{1}{c}{\makecell[c]{Whole image and \\corresponding vessel \\mask as input}} & \makecell[c]{U-Net with EfficientNet-B0  as \\encoder, depth-wise separable \\convolutions as decoder} & \makecell[c]{RIGA, \\IDRiD, PALM} & \makecell[c]{Used clinical prior knowledge \\of the position between the \\OD and the fundus vessels}  \\
    \textbf{MIG}   &   \makecell[c]{coarse to fine \\segmentation}   & \makecell[c]{CENet, which is based on the \\U-Net model, in which the encoder\\ is replaced by ResNet34} & \makecell[c]{ORIGA, \\Drishti-GS1, \\RIMONE\_r3} & \makecell[c]{Used a texture encoder module (contained\\ a dense atrous convolution and a \\residual multi-kernel pooling modules)} \\
    \textbf{EyeStar} &   \makecell[c]{coarse to fine \\segmentation}   & Vision Transformer &\makecell[c]{Drishti-GS,\\ RIM-ONE-r3} & \makecell[c]{Utilized a novel transformer-\\based medical image segmentation \\algorithm}\\
    \textbf{MIAG ULL} &   \makecell[c]{coarse to fine \\segmentation}   & \makecell[c]{PSPNet with ResNet50 as \\base model for both \\coarse and fine segmentation} & \makecell[c]{-}     & \makecell[c]{Three models were respectively\\used to segment the coarse OD mask, \\fine OD mask, and fine OC mask} \\
    \hline
    \hline
    \end{tabular}%
  \label{tab:methods-segmentation}%
  }
\end{table*}%

\textit{Localization of fovea.} 
Table~\ref{tab:task3-summary} shows the methods overview of the 6 teams. We can see that except for the cheeron team, other teams all deal with the localization task as the regression task. The cheeron team utilized YOLOv5 (\cite{YOLOV5-Github-web}) as the object detection model, and the other teams adopted U-Net (\cite{ronneberger2015u}) or HRNet (\cite{9052469}) to achieve the regression prediction. For a detailed description of all methods, please refer to the Appendix.

\begin{table*}[tb]
\renewcommand\arraystretch{4} 
  \centering
  \caption{Methods overview of the 6 participating teams in fovea localization task.}
    \resizebox{1.0\textwidth}{!}{%
    \begin{tabular}{ccccc}
    \hline
    \hline
    \textbf{Team}  & \textbf{Strategy} & \textbf{Architecture} & \textbf{Additional datasets} & \textbf{Highlight} \\
    \hline
    \textbf{MAI}   & \makecell[c]{distance map \\regression} & \makecell[c]{U-Net with\\ EfficientNet-B5\\ as encoder} & -     & \makecell[c]{Utilized test-time training strategy\\ by adding a self-supervised task\\ to update the parameters of the feature\\ extraction module during test-time} \\ 
    \textbf{\makecell[c]{VUNO EYE\\ TEAM}} & \makecell[c]{segmentation and \\offset map\\ regression } & \makecell[c]{U-Net with \\EfficientNet-B0 and \\-B4 as encoder} & private dataset & \makecell[c]{Transformed the location task into segmentation \\and regression tasks to predict the confidence map,\\ x-offset and y-offset map; and  using the prior \\relationship between  fovea and fundus vessel} \\
    \textbf{cheeron} & objection detection & YOLOv5 & \makecell[c]{-} & \multicolumn{1}{c}{\makecell[c]{Directly used the object detection model \\in computer vision, and the ground truth \\of the object box was made according to\\ the fovea coordinates and optic disc diameter.}} \\
    \textbf{EyeStar} & \makecell[c]{Gaussian heatmap \\regression}   & \makecell[c]{HRNet} & -     & \makecell[c]{Fusing the global features of the whole fundus image \\with the local features in the fovea region, \\the feature information is fully used to achieve \\the prediction of the heatmap and coordinate offsets.} \\
    \textbf{MIG}   & \makecell[c]{distance map \\regression} & U-Net & \makecell[c]{-}     & \makecell[c]{Using the prior relationship between fovea and OD,\\  designed Bi-Distance map for predicting\\ the minimum distance to the OD or the fovea}  \\
    \textbf{ALISR} & \makecell[c]{distance map \\regression}  & U-Net  & MESSIDOR & \multicolumn{1}{c}{-} \\    
    \hline
    \hline
    \end{tabular}%
  \label{tab:task3-summary}%
  }
\end{table*}%

\subsection{Strategies for domain adaptation}
In REFUGE2 challenge, the MAI and EyeStar teams designed domain adaptation strategies for the multi-device dataset. Specifically, the MAI team adopted a Test-Time Training (TTT) strategy (\cite{sun2020test}) to ensure their framework can be well generalized to multiple devices in all three tasks. Meanwhile, they also utilized a classical unsupervised domain adaptation (UDA) method (\cite{tsai2018learning}) to deal with the multi-domain data in task 2. The EyeStar team proposed a domain adaptation method by utilizing a external dataset for task 1.

\begin{figure}[tb]
\centering
\includegraphics[width=\linewidth]{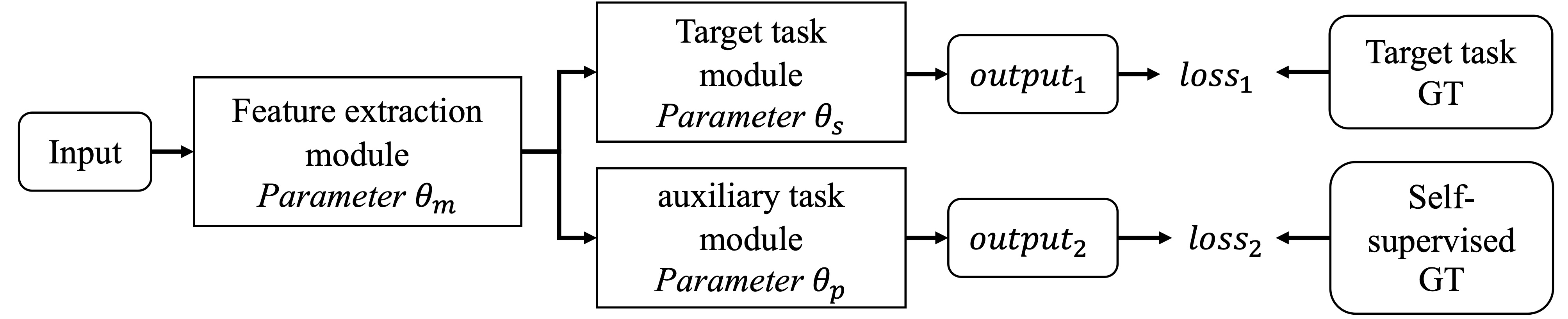}
\caption{Schematic diagram of TTT strategy.}
\label{fig:TTT}
\end{figure}

TTT enforces the framework to optimize itself with test data in the inference stage, which serves as a plug-and-play strategy for model generalization. The key for TTT is to construct a self-supervised auxiliary task for the original baseline. In the training stage, the losses of the target task and the auxiliary task simultaneously supervise the model training and optimize the parameters of $\theta_m$, $\theta_s$, and $\theta_p$ (as shown in Fig.~\ref{fig:TTT}). In the inference stage, the parameters of target task branch and auxiliary task branch are fixed, and the self-supervised auxiliary task is used to fine-tune the public parameters $\theta_m$ of the feature extraction module on the test set, to minimize the $loss_2$. Then, the target task result will be obtained by using fine-tuned public parameters $\theta_m$ and the target task branch parameters $\theta_s$. In MAI team's framework, the auxiliary task consisted of predicting the rotation angle $(0^{\circ}, 90^{\circ}, 180^{\circ}, 270^{\circ})$.

For task 2, the MAI team adopted a classical UDA strategy (\cite{tsai2018learning}). They employed an adversarial training strategy to make the discriminator unable to distinguish from which dataset the prediction comes from, i.e., to force the feature representation of the data from the target domain to be close to the source domain. In their experiments, the domain of the REFUGE2 training set was the source domain, and that of the online set was the target domain.

\begin{figure}[tbh]
\centering
\includegraphics[width=\linewidth]{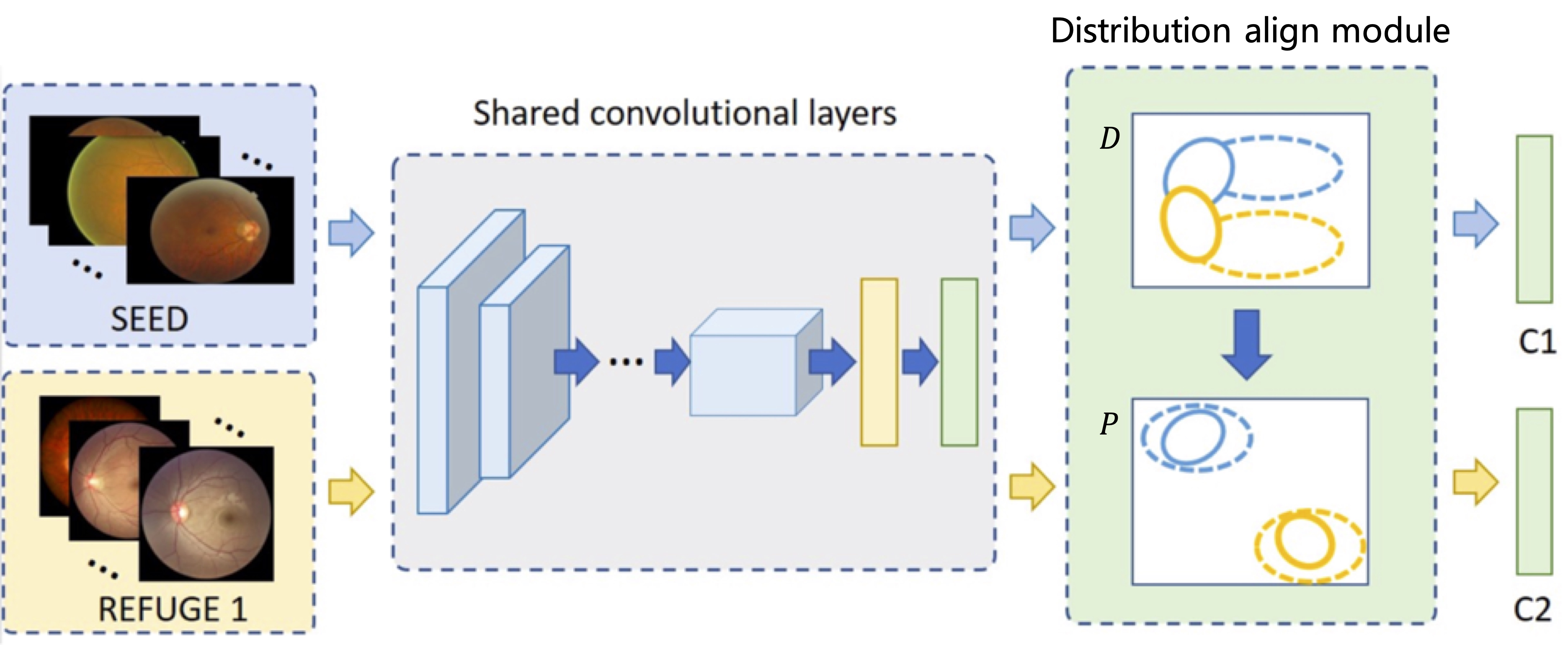}
\caption{The framework of the EyeStar team in Task 1. In distributions align module, yellow represents the sample with label 0 and blue represents the sample with label 1; the solid line represents the sample in REFUGE1, and the dotted line represents the sample in SEED.}
\label{fig:eyestar}
\end{figure}

The EyeStar team used a novel domain adaptation method to transfer the knowledge from their large private dataset, called the Singapore Epidemiology of Eye Disease (SEED) dataset (\cite{guidoboni2020mechanism}), to improve the performance on the REFUGE2 dataset. Specifically, the SEED dataset was used as the source domain and the REFUGE2 training dataset (REFUGE1) was used as the target domain. They proposed to align distributions of the two domains progressively by utilizing the label information. First, two classifiers to predict glaucoma from the source and the target datasets were trained, and these two classifiers shared the feature extraction layers. Then, they defined the sample distribution in the feature space as a conditional distribution $D$ and the sample distribution based the label information as an ideal distribution $P$, during training, $D$ was progressively transferred to be similar to $P$ (as shown in Fig.~\ref{fig:eyestar}), i.e., the sample distributions of the two datasets were aligned in the shared feature space. 

\section{Results}
In this section, the results of the teams in three tasks on the online and onsite set are introduced. In REFUGE2 challenge, the online set with 400 CFPs acquired using a KOWA device, and the onsite set with 400 CFPs acquired using TOPCON camera are used to evaluated the models during preliminary and final rounds, respectively. Since each team can adjust the hyperparameters and the structure of the model according to the leaderboard during the preliminary round, the online set can be regarded as participating in the model training, while the onsite set is the real test set.

\textit{Classification of Clinical Glaucoma}.  
\begin{table}[tbh]
\small
  \centering
  \caption{Results in terms of AUC of 7 teams in the glaucoma classification task. 95\% CI calculated using Delong’s method (\cite{delong1988comparing}).The rankings are among these 7 teams. The red upward or the blue downward arrows and the numbers represent the numbers of places the team has moved up or down on the onsite set compared to the online set.}     
    \begin{tabular}{cccc}
    \hline
    \textbf{Team Name} & 
    \textbf{\makecell[c]{Online\\(95\% CI)}} & 
    \textbf{\makecell[c]{Onsite\\(95\% CI)}} & \textbf{\makecell[c]{Onsite\\ Rank}}\\
    \hline
    \textbf{\makecell[c]{VUNO EYE\\ TEAM}} & \makecell[c]{0.983\\(0.972-0.991)} &  \makecell[c]{0.883\\(0.844-0.919)} & 1\textcolor{red}{$\uparrow$}\textcolor{red}{(1)}\\
    \textbf{MIG}   & \makecell[c]{0.943\\(0.916-0.963)} & \makecell[c]{0.876\\(0.832-0.916)} & 2\textcolor{red}{$\uparrow$}\textcolor{red}{(4)}\\
    \textbf{MAI}   & \makecell[c]{0.9840\\(0.974-0.993)} &  \makecell[c]{0.861\\(0.816-0.904)} & 3\textcolor{blue}{$\downarrow$}\textcolor{blue}{(2)}\\
    \textbf{cheeron} & \makecell[c]{0.980\\(0.970-0.988)} &  \makecell[c]{0.856\\(0.811-0.900)} & 4\textcolor{blue}{$\downarrow$}\textcolor{blue}{(1)}\\
    \textbf{MIAG ULL} & \makecell[c]{0.939\\(0.912-0.960)} &  \makecell[c]{0.847\\(0.798-0.893)} & 5\textcolor{red}{$\uparrow$}\textcolor{red}{(2)}\\
    \textbf{ALISR} & \makecell[c]{0.947\\(0.924-0.966)} &  \makecell[c]{0.844\\(0.796-0.882)} & 6\textcolor{blue}{$\downarrow$}\textcolor{blue}{(2)}\\   
    \textbf{EyeStar} &\makecell[c]{ 0.944\\(0.915-0.969)} &  \makecell[c]{0.820\\(0.766-0.868)} & 7\textcolor{blue}{$\downarrow$}\textcolor{blue}{(2)}\\
    \textbf{vCDR-based} & \makecell[c]{0.8817\\(0.840-0.917)} &  \makecell[c]{0.7571\\(0.693-0.815)} & - \\
    \hline
    \end{tabular}%
  \label{tab:task1}%
 
\end{table}%
Table~\ref{tab:task1} shows the AUC values of the 7 teams on the online and onsite sets. The two-sided 95\% confidence intervals (CIs) here are calculated using Delong’s method (\cite{delong1988comparing, singh2021impact}). We can observe that on the online set, the AUCs for all teams are above 0.93. The performance on the onsite set, with the best performing VUNO EYE TEAM achieves the AUC of 0.883. In our view, there may be two reasons for these results. As mentioned at the beginning, teams can tune their models to the best on the online set via multiple submissions, so the classification performances on the online set are better than those on the onsite set. Moreover, the characteristics of glaucoma samples in the onsite set may bias from those in the online set, so they are not well captured by the models trained and tuned by the training and online sets. 
Here, we adopt DeLong’s test to calculate the statistical significantly difference of the AUC performance (\cite{delong1988comparing, 6851192}). For the online and onsite sets, the results of the 1st teams are not statistical significantly different with respect to those of the 2nd and 3rd teams. Specifically, the $p$ value of significance between the AUCs of the 1st (MAI) and the 2nd team (VUNO EYE TEAM), and 3rd team (cheeron) on online set are 0.974 and 0.850, respectively. The $p$ values are 0.614 and 0.134 for the significance between the AUCs of the 1st (VUNO EYE TEAM) and the 2nd team (MIG), and 3rd team (MAI) on onsite set. 

For comparison, we also include the results, based on using the vCRD values of the ground truth as a likelihood for glaucoma classification, with the AUCs of 0.8817 and 0.7571 on the online and onsite sets, respectively. The trend of vCDR-based classification results on the two sets is consistent with that of the automatic deep learning-based methods. This suggests that the association between glaucoma and vCDR on the onsite dataset is weaker compared to that on the online dataset. This can occur when there are glaucoma cases that do not exhibit optic cup enlargement and some normal cases that have physiological cup enlargements. This indicates that the vCDR alone cannot be used as an independent indicator for automatic discrimination of glaucoma. In our experiments, the glaucoma classification results obtained by the 1st teams on both online and the onsite sets was statistical significantly different with respect to those obtained by the vCDR-based method ($p_{value} = 1.478 \times 10^{-5}$ on online set, and 
$p_{value} = 1.438 \times 10^{-5}$ on onsite set).




\begin{table}[tb]
  \small
  \centering
  \caption{Evaluation in terms of OC Dice, OD Dice and vCDR MAE of the results of 6 teams in the segmentation of optic disc and cup task on the online dataset.}  
    \begin{tabular}{cccc}
    \hline
    \textbf{Team Name} & \textbf{\makecell[c]{OC Dice\\(95\%CI)}} & \textbf{\makecell[c]{OD Dice\\(95\%CI)}} & \textbf{\makecell[c]{vCDR\\ MAE\\(95\%CI)}} \\
    \hline
    \textbf{MAI}   & \makecell[c]{0.880\\(0.873-0.888)} & \makecell[c]{0.966\\(0.965-0.968)} & \makecell[c]{0.037\\(0.033-0.040)} \\
    \textbf{cheeron} & \makecell[c]{0.874\\(0.866-0.882)} & \makecell[c]{0.965\\(0.963-0.967)} & \makecell[c]{0.038\\(0.035-0.042)} \\
    \textbf{\makecell[c]{VUNO EYE \\TEAM}} & \makecell[c]{0.870\\(0.862-0.877)} & \makecell[c]{0.966\\(0.964-0.968)} & \makecell[c]{0.040\\(0.036-0.043)} \\
    \textbf{EyeStar} & \makecell[c]{0.873\\(0.865-0.880)} & \makecell[c]{0.961\\(0.958-0.963)} & \makecell[c]{0.039\\(0.036-0.042)} \\
    \textbf{MIAG ULL} & \makecell[c]{0.854\\(0.845-0.863)} & \makecell[c]{0.934\\(0.932-0.937)} & \makecell[c]{0.044\\(0.041-0.048)} \\
    \textbf{MIG}   & \makecell[c]{0.825\\(0.834-0.853)} & \makecell[c]{0.959\\(0.956-0.961)} & \makecell[c]{0.060\\(0.044-0.052)} \\
    \hline
    \end{tabular}%
  \label{tab:task2-online}%

\end{table}%

\begin{table}[tb]
  \centering
  \caption{Evaluation in terms of OC Dice, OD Dice and vCDR MAE for the 6 teams on the onsite set. The rankings in the table are among these teams. The red upward or the blue downward arrows and the numbers represent the numbers of places the team has moved up or down on the onsite set compared to the online set.}     

     \begin{tabular}{ccccc}
    \hline
    \textbf{Team Name} & \textbf{\makecell[c]{OC Dice\\(95\%CI)}} & \textbf{\makecell[c]{OD Dice\\(95\%CI)}} & \textbf{\makecell[c]{vCDR\\ MAE\\(95\%CI)}} & \textbf{Rank}\\
    \hline
    \textbf{cheeron} & \makecell[c]{0.865\\(0.854-0.875)} & \makecell[c]{0.961\\(0.958-0.964)} & \makecell[c]{0.055\\(0.050-0.060)} & 1\textcolor{red}{$\uparrow$}\textcolor{red}{(1)}\\
    \textbf{MAI}   & \makecell[c]{0.854\\(0.843-0.865)} & \makecell[c]{0.960\\(0.957-0.963)} & \makecell[c]{0.060\\(0.055-0.066)} & 2\textcolor{blue}{$\downarrow$}\textcolor{blue}{(1)}\\
    \textbf{\makecell[c]{VUNO EYE \\ TEAM}} & \makecell[c]{0.845\\(0.836-0.854)} & \makecell[c]{0.960\\(0.956-0.964)} & \makecell[c]{0.058\\(0.053-0.063)} & 2\textcolor{red}{$\uparrow$}\textcolor{red}{(1)}\\
    \textbf{MIG}   & \makecell[c]{0.846\\(0.834-0.858)} & \makecell[c]{0.949\\(0.944-0.953)} & \makecell[c]{0.055\\(0.050-0.060)} & 4\textcolor{red}{$\uparrow$}\textcolor{red}{(2)}\\
    \textbf{EyeStar} & \makecell[c]{0.831\\(0.822-0.840)} & \makecell[c]{0.939\\(0.935-0.942)} & \makecell[c]{0.054\\(0.050-0.058)} & 5\textcolor{blue}{$\downarrow$}\textcolor{blue}{(1)}\\
    \textbf{MIAG ULL} & \makecell[c]{0.851\\(0.838-0.865)} & \makecell[c]{0.918\\(0.905-0.931)} & \makecell[c]{0.064\\(0.056-0.071)} & 6\textcolor{blue}{$\downarrow$}\textcolor{blue}{(1)}\\
    \hline
    \end{tabular}%
  \label{tab:task2-onsite}%

\end{table}%

\textit{Segmentation of Optic Disc and Cup}. Table~\ref{tab:task2-online} and 
Table~\ref{tab:task2-onsite} summarize the OD and OC Dice and vCDR MAE metrics of each team on the online and onsite sets. From the tables, we can see that the performances of OC and OD segmentation models on these two sets are close, which indicates that these segmentation models have better generalization ability. Meanwhile, we can also find that the Dice values of OD are greater than those of OC, which is mainly because OC is smaller and more difficult to segment. Specifically, on the online set, MAI reaches the best segmentation performance with OD Dice of 0.966, OC Dice of 0.880, and vCDR MAE of 0.037. To compare the statistical significance of the differences in the metric values of the top three teams, we adopt Mann-Whitney U hypothesis test with $\alpha=0.05, n_1=n_2=400$. For OD segmentation, compared with 2nd (cheeron) and 3rd (VUNO EYE TEAM) teams, MAI was not statistical significantly different with respect to them (cheeron $p_{value}=0.8037$, VUNO EYE TEAM $p_{value}=0.3715$). No statistical significant different results also occurred among the 1st team respect to the 2nd and 3rd teams for OC segmentation (cheeron $p_{value}=0.1821$, VUNO EYE TEAM $p_{value}=0.0889$) and vCDR evaluation (cheeron $p_{value}=0.2521$, VUNO EYE TEAM $p_{value}=0.1814$). On the onsite set, the 1st place is cheeron with OD Dice of 0.961, OC Dice of 0.865, and vCDR MAE of 0.055. 
For OD segmentation, compared with MAI and VUNO EYE TEAM-the 2nd and 3rd teams, respectively-the differences are not significant (MAI $p_{value}=0.5879$, VUNO EYE TEAM $p_{value}=0.5075$). For OC segmentation, the differences in the OC Dice values achieved by cheeron is statistically significant with respect to VUNO EYE TEAM ($p_{value}=1.0047 \times 10^{-6}$), except to MAI ($p_{value}=0.1057$).  For vCDR estimation, cheeron is with no significant differences with respect to the MAI and VUNO EYE TEAM (MAI $p_{value}=0.0715$, VUNO EYE TEAM $p_{value}=0.0795$).

\begin{figure}[tb]
    \centering    \includegraphics[width=0.75\linewidth]{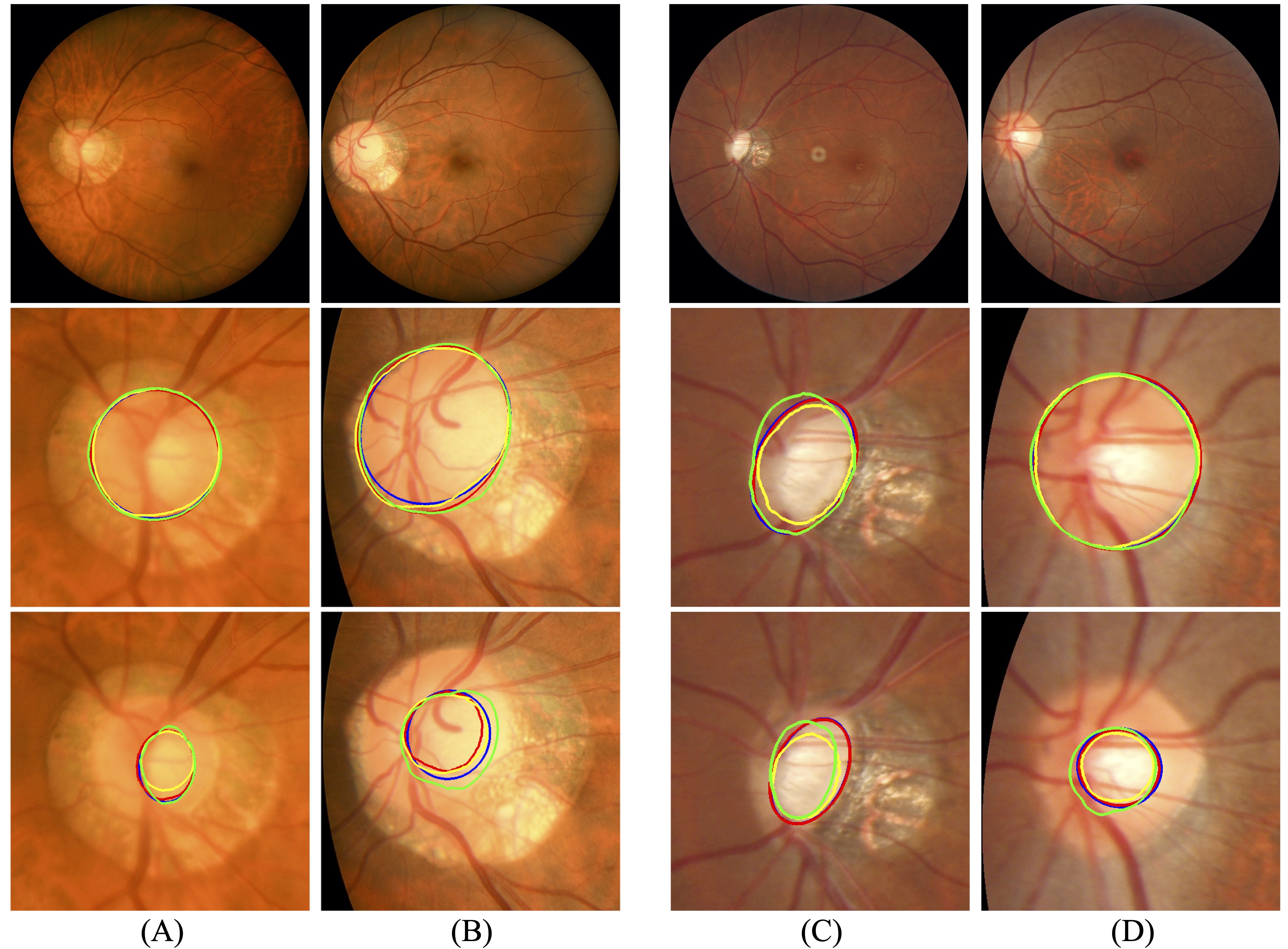}
    \caption{OD/OC segmentation results on the online and onsite datasets of the top 3 teams. (A)-(B) Glaucoma and non-glaucoma samples in the online dataset, Green: Ground truth, Blue:MAI, Red: cheeron, Yellow: VUNO EYE TEAM. (C)-(D) glaucoma and non-glaucoma samples in the onsite dataset, Green: Ground truth, Blue: cheeron, Red: MAI, Yellow: VUNO EYE TEAM.}
    \label{fig:seg_results}
\end{figure}

Fig.~\ref{fig:seg_results} shows the contours of the OD and OC segmentation results on the online and onsite sets of the top 3 teams, respectively. Figs.~\ref{fig:seg_results}(A) and (C) are glaucoma samples in the online and onsite sets, and Figs.~\ref{fig:seg_results}(B) and (D) are non-glaucoma samples. In Figs.~\ref{fig:seg_results}(A) and (B), green, blue, red and yellow lines respectively present the ground truth, and the segmentation results of MAI, cheeron, and VUNO EYE TEAM. Similarly, 
In Figs.~\ref{fig:seg_results}(C) and (D), the four color lines respectively present the ground truth, the results of cheeron, MAI, and VUNO EYE TEAM. From the figure, we can see that the segmentation results on both glaucoma and non-glaucoma images can cover the target region.

\begin{table}[tb]
\small
  \centering
  \caption{Results in terms of AED of 6 teams in the fovea localization task on both online and onsite datasets. The rankings in the table are among these teams.}
    
    \begin{tabular}{cccc}
    \hline
    \textbf{Team Name} & \textbf{\makecell[c]{Online\\(95\% CI)}} & \textbf{\makecell[c]{Onsite\\(95\% CI)}} & \textbf{\makecell[c]{Onsite\\ Rank}}\\
    \hline
    \textbf{MAI}   &  \makecell[c]{8.412\\(7.670-9.155)} &  \makecell[c]{21.841\\(17.863-25.818)} & 1\textcolor{gray}{(-)}\\
    \textbf{\makecell[c]{VUNO EYE \\TEAM}} &  \makecell[c]{8.727\\(7.864-9.590)}  &  \makecell[c]{27.456\\(20.496-34.415)} & 2\textcolor{gray}{(-)}\\
    \textbf{cheeron} &  \makecell[c]{10.086\\(9.195-10.976)}  &  \makecell[c]{28.344\\(20.281-36.407)} & 3\textcolor{gray}{(-)}\\
    \textbf{EyeStar} &  \makecell[c]{10.096\\(8.866-11.327)} &  \makecell[c]{43.982\\(33.628-54.336)} & 4\textcolor{gray}{(-)}\\
    \textbf{MIG}   &  \makecell[c]{35.741\\(25.808-31.973)}  &  \makecell[c]{105.812\\(85.028-126.596)} & 5\textcolor{gray}{(-)}\\
    \textbf{ALISR} &  \makecell[c]{111.239\\(87.135-135.344)}  &  \makecell[c]{173.405\\(146.393-200.418)} & 6\textcolor{gray}{(-)}\\
    \hline
    \end{tabular}%
  \label{tab:results-localization}%

\end{table}%

\textit{Localization of Fovea}. Table~\ref{tab:results-localization} summaries the results of the 6 teams in the fovea localization task on the online and onsite sets. As can be seen from the table, the effects of the localization models on the onsite set are worse than those on the online set, which may be caused by the overfitting of the models tuned with the online set, or by the different data domains. Specifically, on the online set, the first 3 teams are MAI, VUNO EYE TEAM, and cheeron with the AED of 8.412, 8.727, and 10.086 pixels, respectively. Compared to VUNO EYE TEAM and cheeron, the performance of MAI is only statistical significantly different with respect to cheeron (VUNO EYE TEAM $p_{value}=0.6356$, cheeron $p_{value}=0.0003$). On the onsite set, the first 3 teams are also MAI, VUNO EYE TEAM, and cheeron with AED of 21.841, 27.456, and 28.344 pixels, respectively. The performance of MAI is not statistical significantly different compared to VUNO EYE TEAM and cheeron (VUNO EYE TEAM $p_{value}=0.3764$, cheeron $p_{value}=0.3283$).


\begin{figure}[tb]
    \centering    \includegraphics[width=0.75\linewidth]{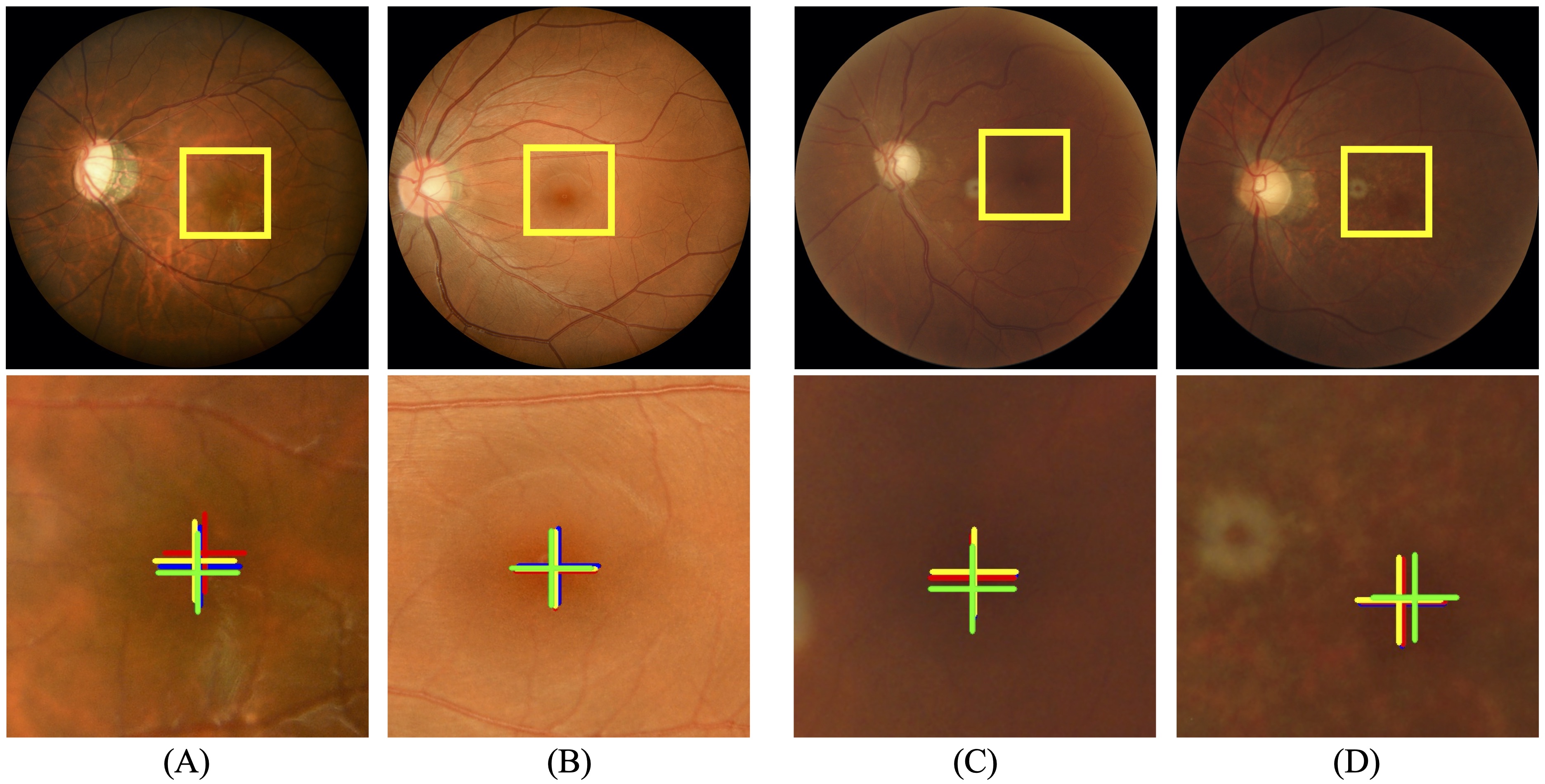}
    \caption{Fovea localization results of the top 3 teams on the enlarged display. (A) Glaucoma sample in the online dataset, (B) non-glaucoma sample in the online dataset; (C) glaucoma sample in the onsite dataset, (D) non-glaucoma sample in the onsite dataset. Green: Ground truth, Blue: MAI, Red: VUNO EYE TEAM, Yellow: cheeron.}
    \label{fig:fovea_loc}
\end{figure}

Fig.~\ref{fig:fovea_loc} shows the fovea localization results of the top 3 teams on the online and onsite sets, respectively. In the figure, the first row shows the original images, and the second row shows the corresponding fovea localization results with the cross-marks. Similarly, we show the results on the glaucoma and non-glaucoma samples. It can be seen that the localization results of the top 3 teams on the online and onsite sets are near the ground-truth for both glaucoma and non-glaucoma samples.

\section{Discussion}
In this section, the methodological findings are discussed in Section~\ref{findings} after analyzing the challenge results. The impact of glaucoma on fundus structure analysis and the impact of image quality on glaucoma assessment and fundus structure analysis are discussed in Section~\ref{effects}. In Section~\ref{correlation}, we discuss the correlation among the results obtained by the models designed by the participating teams, and we also discuss the ensemble performances of these models. Finally, the clinical implications, and limitation and future work of our challenge are discussed in Sections~\ref{cinical implications} and~\ref{future}, respectively.

\subsection{Methodological findings}
\label{findings}
\textit{Strategies for domain adaptation.} We examine the released multi-domain dataset, and observe that there are indeed differences in the image distribution (Fig.~\ref{fig:data_distribution}) of the CFPs collected by different devices. Such data are especially valuable for studying the methods of domain adaptation. In REFUGE2 challenge, the MAI and EyeStar teams consider the domain adaptation strategies. From the results of each subtask, the MAI team always ranks 1st on the online set. This shows that the TTT method is very effective when there is a domain bias between the training data and the test data. Since our challenge required that the weights of the model could not be updated in the final round, the weights of the feature extraction module designed by the MAI team were not updated with the onsite set, i.e., the model did not play the role of domain adaptation, resulting in a drop in the final ranking. In task 1, the VUNO EYE TEAM and MIG teams do better on the onsite set, and we can see that these teams utilize the additional training data. This reflects another approach to dealing with the problem of different data domains, i.e., using domain-diverse training data. As can be seen in Table~\ref{tab:datasets}, part of the samples in the additional dataset ACRIMA used by the MIG team were collected with the Topcon equipment, which is consistent with the device used in the onsite set. Although the EyeStar team used an additional dataset and domain adaptation strategy, they were aligning the feature space of the training set of REFUGE2 and the private dataset SEED by considering the data domain of the REFUGE2 training set as the target domain. However, the data domains of the online and onsite sets to be tested in the challenge are different from the training domain, so the model that completed the feature space alignment above did not achieve domain adaptation on the online and onsite sets. Therefore, the team did not perform well in Task 1. Since the method of the EyeStar team needs to align the feature space of different domains using label information, the method is not suitable for domain adaptation tasks where the labels are unknown in the target domains.

\textit{Classification of clinical glaucoma.} Glaucoma will lead to degeneration around the optic disc region, such as vCDR expansion, optic disc bleeding, optic nerve rim notching and other signs (\cite{aung2016asia}). Hence, 6 of 7 teams predicted the glaucoma considering the local optic disc region. Among them, the MIG team utilized not only the local region of the optic disc, but also the whole fundus image. The abundant global and local features can improve the model's performance. The network architectures used in the glaucoma classification task included EfficientNet, variants of ResNet, VGG, and DenseNet, which are the most common and conventional neural networks for image classification. Among them, the ALISR team adopted a cross-stage partial strategy to alleviate the problem of heavy inference computation. The solution to alleviate computational burden in the application of AI technology is a valuable topic (\cite{wang2020cspnet}). From Table~\ref{tab:task1}, we can see that the glaucoma classification results obtained from CFP with AI are superior to those obtained using vCDR value directly, which is the key indicator in glaucoma screening, indicating that deep learning methods may advance the state of the art in glaucoma screening.


\textit{Segmentation of optic disc and cup.} From the aspect of the network architectures, U-Net and its variants remain popular for the segmentation task. Notably, the EyeStar team used vision transformer technology, which emerged as a competitive alternative to convolutional neural networks that are the current state of the art in computer vision (\cite{khan2021transformers}). However, this technique has high requirements on the training data size and computational resources. 
In addition, we have compared the initial annotations, which were delineated by different glaucoma specialists, with the ground truth of the OD and OC segmentation task in the onsite set. The results show that the best performance of the manual delineation is $Dice_{OC}=0.865, Dice_{OD}=0.952, MAE_{vCDR}=0.049$, the worst one is $Dice_{OC}=0.742, Dice_{OD}=0.817, MAE_{vCDR}=0.084$.As can be seen from Table~\ref{tab:task2-onsite}, the results obtained automatically are all better than the worst one by the manual annotator. Meanwhile, the performance of the cheeron team (Rank 1) is closest to the best manual delineation. This indicates that the automated segmentation methods can achieve similar or even better performance than manual annotations, which can serve to assist the analysis of the OD and OC in clinical practice.

\textit{Localization of fovea.} In this task, we observe that the proposed solutions mostly transform the localization task to segmentation, distance map or heatmap regression, and object detection tasks. The corresponding ground truths are also converted into the corresponding forms. In the distance map or heatmap regression and segmentation tasks, common network frameworks such as U-Net were mainly used. For the object detection task, cheeron team utilized YOLOv5, the latest version of the prominent YOLO series. It can be observed that the models typical of the computer vision field are successfully applied in the medical imaging domain. Similar to the segmentation task, we also calculated the differences between the manual annotations and the ground truth for the fovea localization. The AED of the manual annotation result ranged from 22.94 - 27.41 pixels. The average AED across the annotators is 24.96 pixels. As seen from Table~\ref{tab:results-localization}, only the 1st team MAI outperformed all the manually annotated results, and is hence better than the best human annotator. This is mainly due to the fact that the localization task is harder than the segmentation task. Still, the performances of the VUNO EYE TEAM and cheeron teams are close to the worst manual labelling result. This shows that it is possible for automated methods to achieve similar results to manual annotation in the fovea localization task, which plays an important role in the clinical analysis of macular region.

\textit{Clinical prior knowledge.} In addition to the solutions in the classification task emphasizing the clinical prior information of the OD region, the VUNO EYE TEAM considered the anatomical relationship between OD and blood vessels, and between fovea and blood vessels when building the model in the OD/OC segmentation and fovea localization tasks. And, in the localization task, the MIG team used the relationship between the fovea and the OD. Furthermore, in addition to the prior information on the location of fundus structures, the VUNO EYE TEAM (1st in task 1) used the prior information of lesion in the glaucoma classification task. Specifically, they used private datasets with lesion labels that could train the model to extract the lesion features, where the lesion labels such as OD hemorrhage, retinal nerve fiber layer defects, and glaucomatous OD changes are very meaningful for glaucoma classification. The utilization of the clinical prior knowledge shows to be improving results in the challenge. This indicates that clinical prior knowledge is of great significance in computer-aided disease diagnosis and medical image analysis.


\subsection{Effects of disease situation and image quality}
\label{effects}

In this section, we discuss the impact of image quality on glaucoma assessment and fundus structure analysis, as well as the impact of glaucoma on fundus structure analysis.

\begin{figure}[tb]
    \centering    \includegraphics[width=0.75\linewidth]{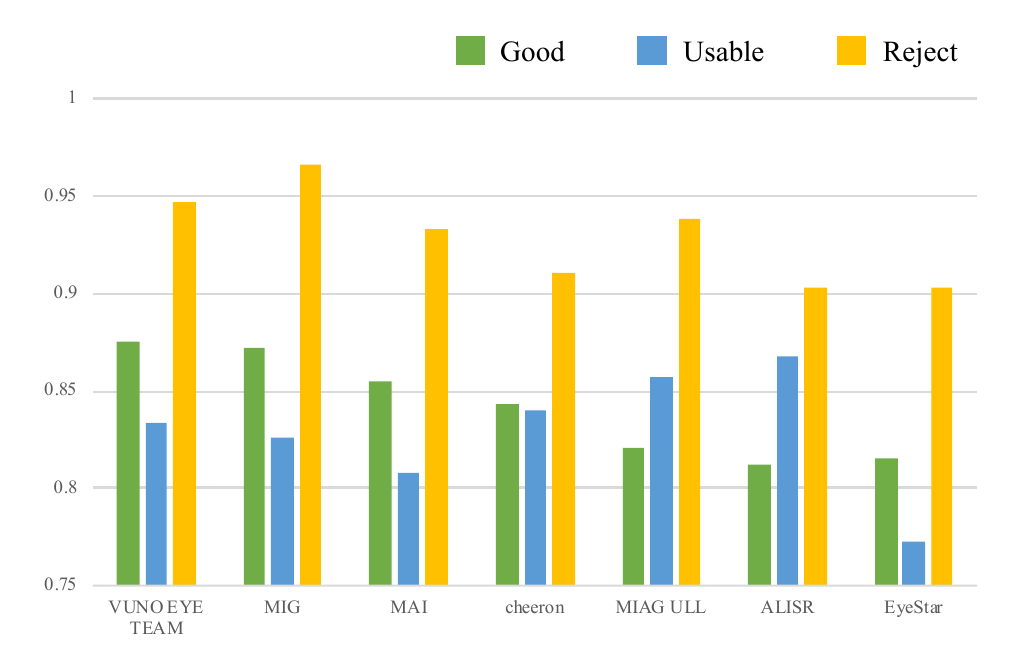}
\caption{Classification results stratified by data image quality.}
    \label{fig:task1-dif-property}
\end{figure}

\begin{figure}[tb]
    \centering    \includegraphics[width=1\linewidth]{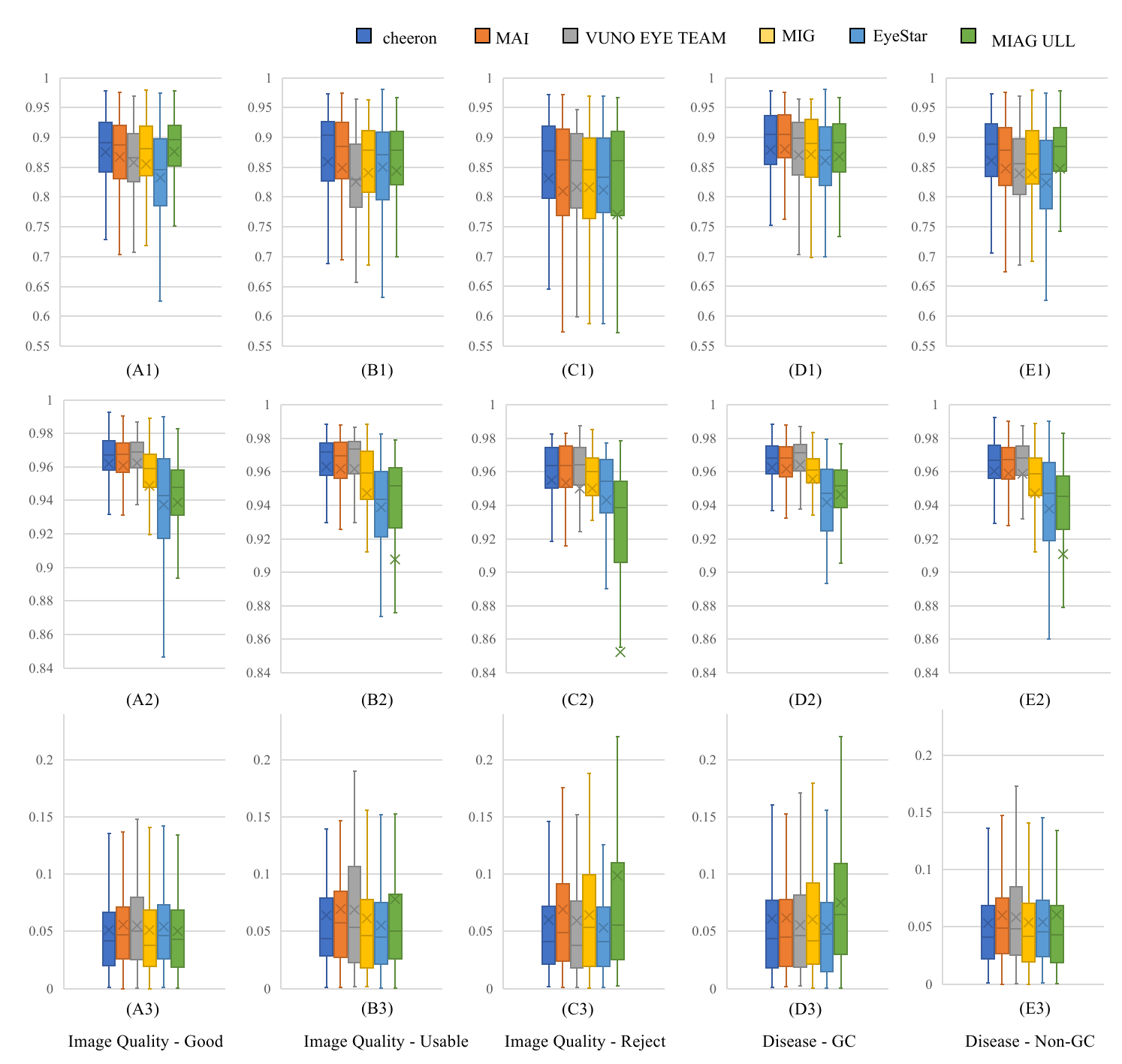}
    \caption{OD/OC segmentation results stratified by data image quality and disease situations. The first to third rows correspond to the OC Dice, the OD Dice, and the vCDR MAE metrics, respectively. The first to third columns correspond to samples with ‘good', ‘usable', and ‘reject' image quality, respectively, and the fourth to fifth columns correspond to glaucoma samples and non-glaucoma samples, respectively.}
    \label{fig:task2-dif_property}
\end{figure}

\begin{figure}[tb]
    \centering
    \includegraphics[width=1\linewidth]{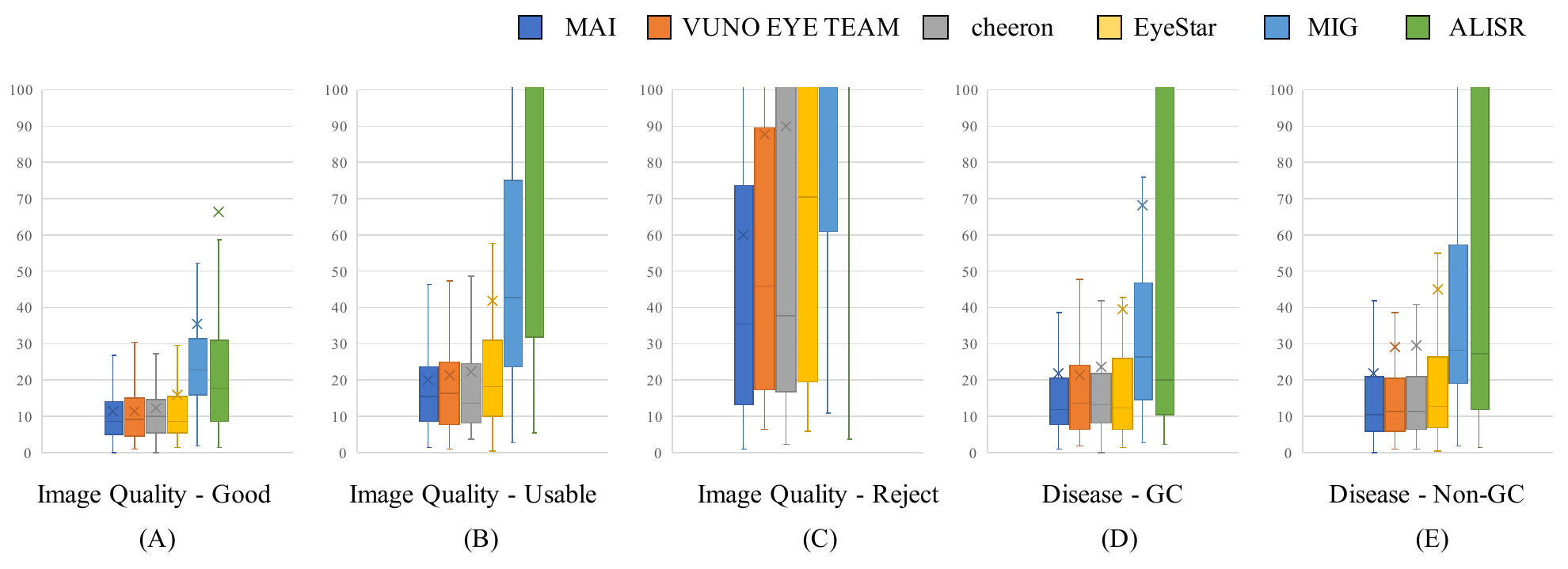}
    \caption{Fovea localization results stratified by data image quality and disease situations. (A)-(C) correspond to samples with ‘good', ‘usable', and ‘reject' image quality, and (D)-(E) correspond to glaucoma samples and non-glaucoma samples.}
    \label{fig:task3-dif_property}
\end{figure}

\textit{Fundus image quality.} Among the CFPs in the REFUGE2 onsite set, 267 images are evaluated as ‘good', 57 images as ‘usable', and the remaining 76 images as ‘reject' according to the retinal image quality assessment method (\cite{fu2019evaluation}), which considers four quality indicators, including blurring, uneven illumination, low-contrast, and artifacts. It should be noted that in the study of Fu et al, CFPs with OD or macular region in low image quality are considered as ‘reject’, so the CFPs which have macular region blurred, not visible, or with artificial artifacts in the dataset are identified as ‘reject’. These quality issues do not involve the OD area, and these images may be present during glaucoma screening in the clinic, so they are included in our dataset. 
Figs.~\ref{fig:task1-dif-property},~\ref{fig:task2-dif_property}(A)-(C), and~\ref{fig:task3-dif_property}(A)-(C) show the evaluation of the predicted results obtained for the three tasks on the CFPs with different qualities, respectively. 

\textit{Effects of image quality on glaucoma classification.} Fig.~\ref{fig:task1-dif-property} shows the classification AUCs on the CFPs with ‘good', ‘usable', and ‘reject' qualities on the onsite set. From the figure, the teams have the best classification performance on the CFPs with the ‘reject’ quality. This is possible in our cases because such ‘reject’ images are mainly underexposed, blurred, or have artifacts in the macula or retinal vessels, while the OD region is clear, so the performance of the glaucoma classification methods designed with the OD region as the main attention is not affected. Moreover, even better classification results may be obtained because of the high contrast between the OD region and the background region in the ‘reject’ CFPs. However, it is worth noting that the number of CFPs of different qualities in the onsite set is not evenly distributed, so to obtain more accurate conclusions, we need to expand the ‘reject' CFPs in the set. It can be seen from Fig.~\ref{fig:task1-dif-property} that the classification performance obtained by the teams in the ‘good' quality images follows the same trend as the overall ranking. For the CFPs with ‘usable' quality, the three teams, cheeron, MIAG ULL and ALISR, can obtain the better results. From Table~\ref{tab:classication-methods}, we can see that the inputs of these teams are all the patches of the OD region. Moreover, the cheeron and ALISR teams both used ResNext (\cite{8100117}), whcih has a stronger capability to capture the useful representative features from images. The MIAG ULL team used a simple VGG19 network without additional tricks, which made the model less susceptible to extracting interference features. Similar to the discussion of the ‘reject' images, we need to supplement the onsite set with more ‘usable' images to reach a more accurate conclusion. With the above analysis, we believe that designing different models for different quality images is a strategy that can be applied in clinical scenarios involving variable quality of fundus images.

\textit{Effects of image quality on fundus structure analysis.} Figs.~\ref{fig:task2-dif_property}(A)-(C) and~\ref{fig:task3-dif_property}(A)-(C) show the evaluation stratified by different image qualities of the 6 teams in terms of OC Dice, OD Dice, and vCDR MAE in the segmentation task and the AED in the localization task. 
As we can see from the figures, especially from the results of the first 3 teams, the OD/OC segmentation task is much less affected by image quality variations than the fovea localization task. This is because blurring, exposure, underexposure, or artifacts, that interfere with the observation of the fundus structures, are not severely involved in the OD region
. From Fig.~\ref{fig:task3-dif_property}(A)-(C) we can see that all methods deteriorate the localization results as the image quality gets worse, especially on CFPs with ‘reject' quality, the localization results deteriorate very significantly. Furthermore, as can be seen in Figs.~\ref{fig:task3-dif_property}(B), for the localization task of samples with slightly poorer image quality, the models designed by the top 3 of the 6 teams are more robust than those designed by the bottom 3 teams. Note that for the cross-sectional comparison, the AED value display in the boxplots for the different cases in Fig.~\ref{fig:task3-dif_property} are adjusted to the range from 0 to 100, and the corresponding complete boxplots can be viewed in the Appendix (Figure B.1).

\textit{Effects of glaucoma on fundus structure analysis.} Comparing Fig.~\ref{fig:task2-dif_property} (D1) and (E1), it can be seen that the Dice evaluation of the OC segmentation results obtained by each team on the non-glaucoma samples are slightly lower than those obtained on the glaucoma samples
In OD segmentation task, the models of the first 3 teams performed more consistently in both glaucomatous and non-glaucomatous samples; the latter 3 teams obtained less error in the glaucomatous samples 
. Similar trend can be observed from the localization results in Fig. ~\ref{fig:task3-dif_property}(D) and (E). We believe that this trend occurs in our onsite set, where the results of fundus structure analysis have slightly less error in the glaucomatous samples, mainly because the number (80) of glaucomatous samples in the set is much lower than that (320) of non-glaucomatous samples. We therefore speculate that glaucoma does not interfere with the analysis of fundus structure. 

\subsection{Predictions correlation and ensemble models}
\label{correlation}



\begin{figure}[tb]
    \centering
    \includegraphics[width=1\linewidth]{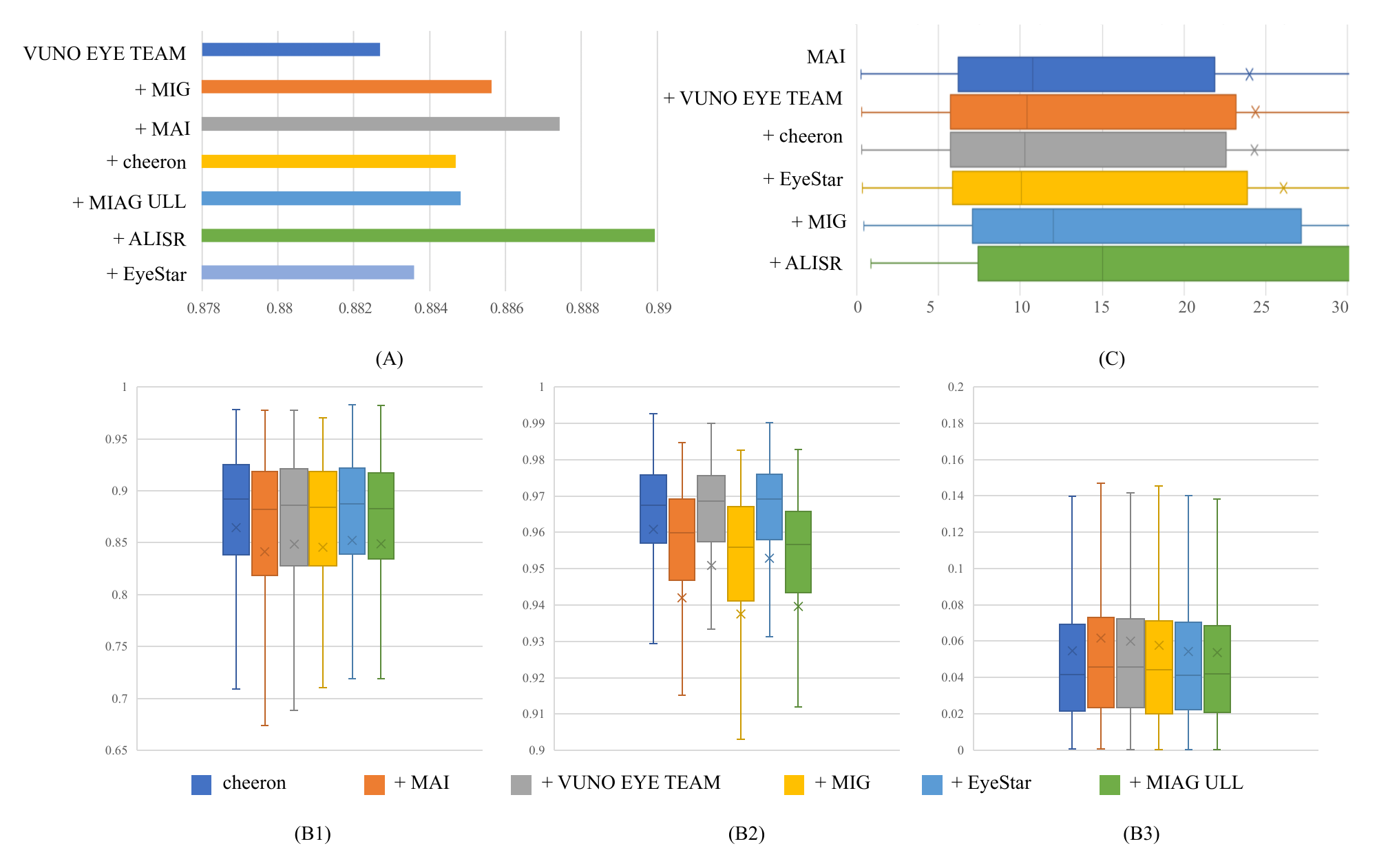}
    \caption{Results of the ensembled models.}
    \label{fig:ensemble}
\end{figure}

\begin{figure}[tb]
    \centering
    \includegraphics[width=1\linewidth]{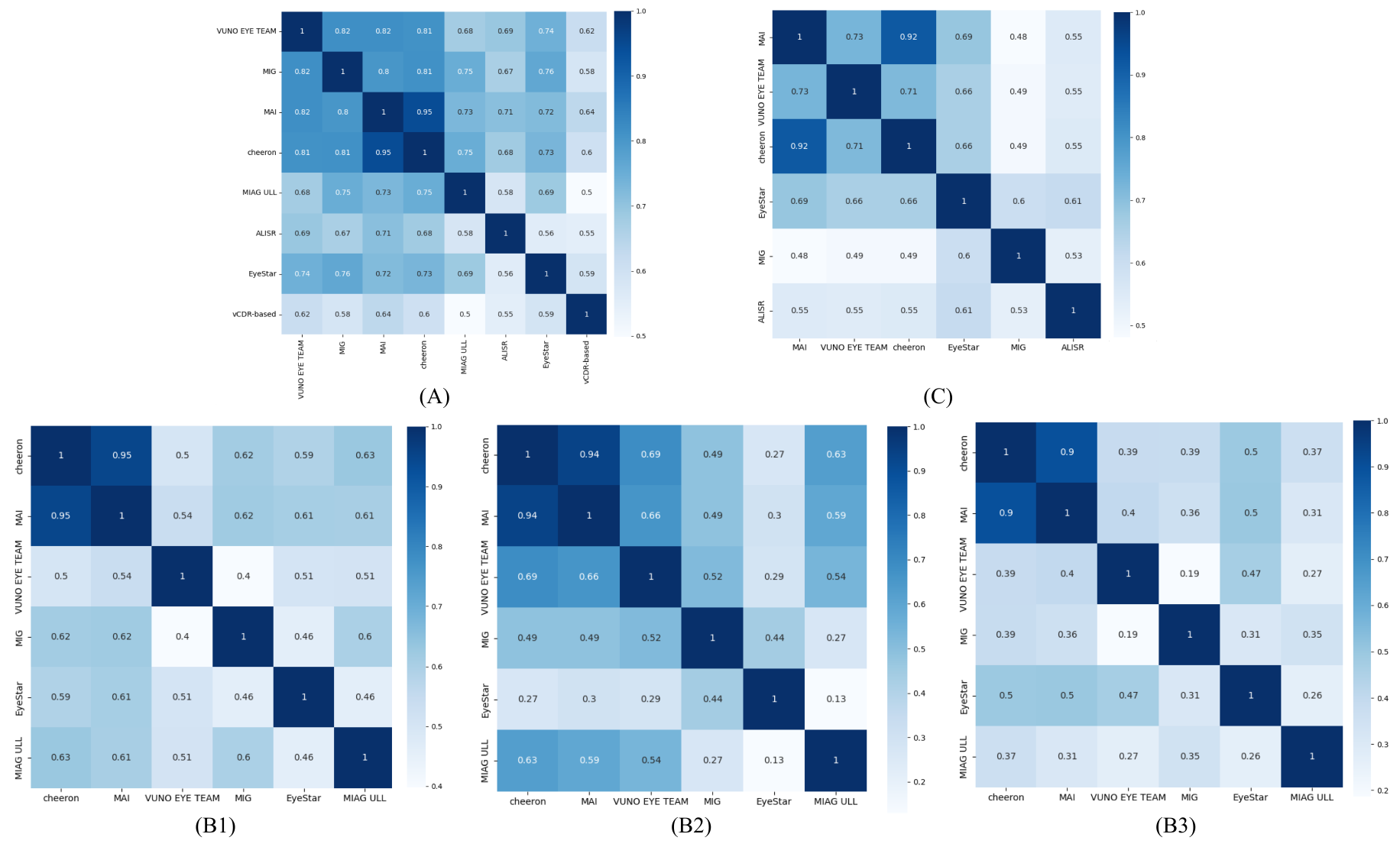}
    \caption{Spearman correction between teams.}
    \label{fig:correction}
\end{figure}

In the three tasks, to discuss the ensemble effect of the different methods, we calculate the Spearman correlation of the results obtained by each method, and composite the predictions of each team in order of ranking (\cite{10.1001/jamanetworkopen.2022.27423}). Fig.~\ref{fig:ensemble} shows the evaluation of the ensemble results. Specifically, Fig.~\ref{fig:ensemble}(A) shows the AUCs for the glaucoma classification after ensemble of the results for each team. We sum and average the classification prediction probabilities obtained by the teams to obtain the ensemble predicted probabilities. The first row in Fig.~\ref{fig:ensemble}(A) represents the result of the VUNO EYE TEAM team with the best performance on the onsite set, and each subsequent row represents the result of the next team ensembled to the model of the previous row. As can be seen from the figure, after adding the results of the 7 teams in turn, the obtained glaucoma classification results are all slightly better than the performance of the 1st team, and the highest effect improvement is obtained when the results of the first 6 teams are ensembled, with an improvement of 0.007 in AUC. As can be seen from the correlation of the results of each team in Task 1 shown in Fig.~\ref{fig:correction}(A), the result of the 6th ranked ALISR team have a relatively low correlation (p=0.58-0.71) with the results of the previous 5 teams, so it provided a positive complement to the predicted probabilities on some samples. However, it is worth noting that the 1st team has achieved good results on the glaucoma classification, as the AUC improved by less than 0.01 after ensembling the results of the other teams.

Figs.~\ref{fig:ensemble}(B1)-(B3) show the results of the ensemble models for the segmentation task in terms of OC Dice, OD Dice, and vCDR MAE metrics, respectively. In our experiments, the ensemble strategy in the segmentation task is performed by majority voting, i.e., the final category of the pixel agrees with more than half of the predicted categories. For example, during the ensemble of 4 teams and if a pixel is predicted as a certain category more than 2 times, the pixel is given the label of that category, otherwise it is considered as background. From the ensemble results, the cheeron team has the best segmentation effect. It can be seen from Figs.~\ref{fig:ensemble} (B1)-(B2) that the performance reductions of the ensemble results with the odd numbered teams are less than those with the even numbered teams. This is mainly because the votes of the predicted OC and OD category for a pixel may be the same when ensemble an even numbered teams, and the pixel will be classified as the background according to the majority voting rules, resulting in the reduction in segmentation accuracy. Note that the scales of the vertical axis in Figs.~\ref{fig:ensemble}(B1) and (B2) are inconsistent in order to see the variation of the results of different ensemble models in terms of every metric clearly. The correlation  (Figs.~\ref{fig:correction}(B1)-(B3)) show that the results of the 1st and 2nd teams in the segmentation task are more correlated, so their ensemble results cannot improve the prediction. Moreover, although the results of the other teams have lower correlation with those of the first 2 teams, they did not positively influence the results.

From Fig.~\ref{fig:ensemble}(C), the result of the MAI team is the best in the fovea localization task. When it is ensembled with the result of the 2nd team, the overall prediction is more deviated. When combined with the result of the 3rd team, the overall bias falls back because the result of the 3rd team correlates more strongly with the MAI result (see Fig.~\ref{fig:correction}(C)). When ensemble with the results of the remaining 3 teams, the biases of the ensemble results become larger because of the poor performances of these 3 teams. Note that for the cross-sectional comparison, the AED values displayed in the boxplot in Fig.~\ref{fig:ensemble}(C) are adjusted to the range from 0 to 30, and the corresponding complete boxplot can be viewed in the Appendix (Figure B.2).

Therefore, when performing ensemble operations, it is important to choose the complementary base models with good performances, and it is better to use an odd numbered base models when using majority voting strategy.

\subsection{Clinical implications of the challenge}
\label{cinical implications}
REFUGE2 challenge has released the first multi-device and multi-task CFP dataset, which can be used to study the application of AI algorithm in glaucoma classification, OD/OC segmentation and fovea localization, and to study the performance of AI algorithms on different imaging qualities and devices. Already, several published studies on domain adaptation have used the REFUGE2 dataset (\cite{li2021few, guo2021cafr}). 
We can find from the challenge results that the current AI-based automatic classification methods can provide more accurate glaucoma recognition results than those obtained by the vCDR-based method commonly used in glaucoma screening. Moreover, in fundus structure analysis such as OC/OD segmentation and fovea localization, AI algorithms can obtain comparable or even better results than manual labeling. We hope that REFUGE2 will follow the success of other challenges within the iChallange series, such as ADAM (\cite{fang2022adam}), PALM (\cite{fu2019palm}), AGE (\cite{fuAGEChallengeAngle2020a}), and bridge the gap between scientific research and clinical application.

\subsection{Limitation and future work}
\label{future}
The limitation of the REFUGE2 challenge is the lack of the associated demographic information of the dataset, such as age distribution and data source (clinic, community). In addition, the dataset were collected exclusively from the Chinese population, limiting the ethnic diversity. 
And, in the evaluation framework, we did not consider the prediction effect of different image quality samples. In future challenges, we will design and document the collection and distribution of datasets in more detail, and consider multi-dimensional evaluation strategies within the evaluation framework. In addition, we will focus on the grading of glaucoma severity as well as multi-modality tasks (\cite{wu2022gamma,fang_dataset_2022}) as the clinical diagnosis of glaucoma involves not only CFP examination, but also OCT, visual field test and other examinations.

\section{Conclusion}

In this paper, we introduce the REFUGE2 challenge, especially the multi-device dataset, provide the solutions adopted by the high-performance teams and the results obtained by these methods, and discuss the findings from the challenge. 
The REFUGE2 dataset is the first open multi-device fundus images focused on glaucoma classification, OD/OC segmentation, and fovea localization, which simulates the clinical scenarios that the CFPs collected by different devices. The current deep-learning methods have good effects on glaucoma assessment and fundus structure analysis in the known data domain, but the prediction effects in the various and unknown data domains need to be improved. Therefore, the domain adaptation method needs to be strengthened, and it will have great potential to be used in clinical practice. 

\section*{Acknowledgments}
\small
This research was supported by the High-level Hospital Construction Project, Zhongshan Ophthalmic Center, Sun Yat-sen University (303020104); the National Natural Science Foundation of China (82070955); the Science and Technology Program of Guangzhou (2021), China.

\renewcommand\thefigure{\thesection.\arabic{figure}}

\section*{Appendix}
\appendix
\section{Challenge Solution Report}
Three clinical tasks are proposed in the REFUGE2 challenge: classification of clinical glaucoma, segmentation of optic disc and cup, and localization of fovea (as shown in Fig. 1 in the paper). This supplementary materials summarize the methods of the VUNO EYE TEAM, cheeron, MAI, MIG, EyeStar, MIAG ULL and ALISR teams, which are the top 10 teams in each task of the overall leaderboard (The remaining 3 teams that met the conditions gave up participating in this paper). In addition, the methods of Pami-G, TeamTiger and CBMIBrand teams with better performance in semi-final leaderboard are also summarized. 

\subsection{Classification of clinical glaucoma}
The computer-aided clinical glaucoma classification is to predict the probability of a given color fundus image belonging to glaucoma. 
Three teams proposed classification methods based on the whole images. The optic disc (OD) region is critical for glaucoma prediction due to the significant variety of the structure and texture in the OD and optic cup (OC) region caused by glaucoma, hence, the remaining teams all considered the local information in the OD region. 

The VUNO team modified the architecture introduced in previous study (\cite{son2020development}), which could predict fifteen ophthalmologic lesions based on the whole color fundus image, with EfficientNet. The specific method was the same as the team used in ADAM Challenge task1 (\cite{fang2022adam}). The TeamTiger team trained Efficientnet family architectures, namely EfficientNet-B4, EfficientNet-B5, EfficientNet-B6, and EfficientNet-B7 to conduct the glaucoma classification. Lastly, the results from these four architectures were ensembled by averaging. The EyeStar also used the whole color fundus images, while they used a novel domain adaptation method to transfer the knowledge from their bigger private dataset to improve the performance on the REFUGE2 training dataset (as introduced in Section \textit{3.2 Strategies for domain adaptation)}.

There were six teams predicting glaucoma based on the local OD region only. They all first segmented OD region coarsely, and cropped the OD patch with designed size, then they adopted different models to predict the glaucoma based on the OD patch. In particular, the cheeron team cropped the OD patches with 3 disc diameter (DD) and adopted Res2NeXt, which is the combination of ResNeXt (\cite{xie2017aggregated}) and Res2Net (\cite{gao2019res2net}), to predict glaucoma probability. As shown in Fig~\ref{fig:cheeron-task1}, to increase the effectiveness of the extracted feature maps, they applied an attention module after the Block 2 to increase the network’s attention to the relevant features and suppress the unnecessary features. Besides, the MAI team also predicted glaucoma using only the local OD region, as shown in Fig.~\ref{fig:MAI-task1}. They first acquired the coarse OD segmentation mask with a standard U-Net (\cite{ronneberger2015u}). Then, in their classification stage, the region of interests (ROIs) centered on the OD with 2.5 DD size as the length were cropped and resized to 256×256, and ResNet50 (\cite{7780459}) was chosen as the baseline for classification. To increase the generalization capability of the model for different devices, they proposed to integrate the Test-Time Training strategy (TTT) as introduced in Section \textit{3.2 Strategies for domain adaptation} in the paper. 

\setcounter{figure}{0}   
\begin{figure}[tb]
\centering
\includegraphics[width=\linewidth]{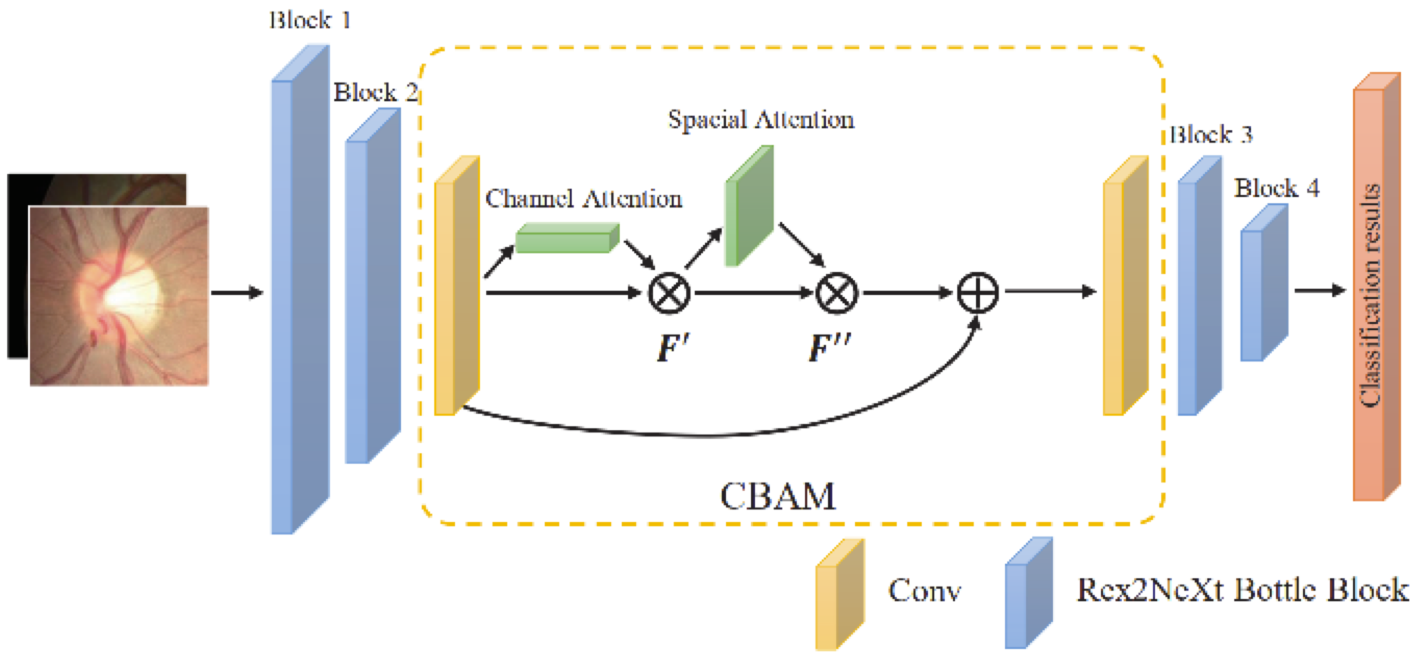}
\caption{The pipeline of the method of the cheeron team in Task1.}
\label{fig:cheeron-task1}
\end{figure}

\begin{figure}[tb]
\centering
\includegraphics[width=\linewidth]{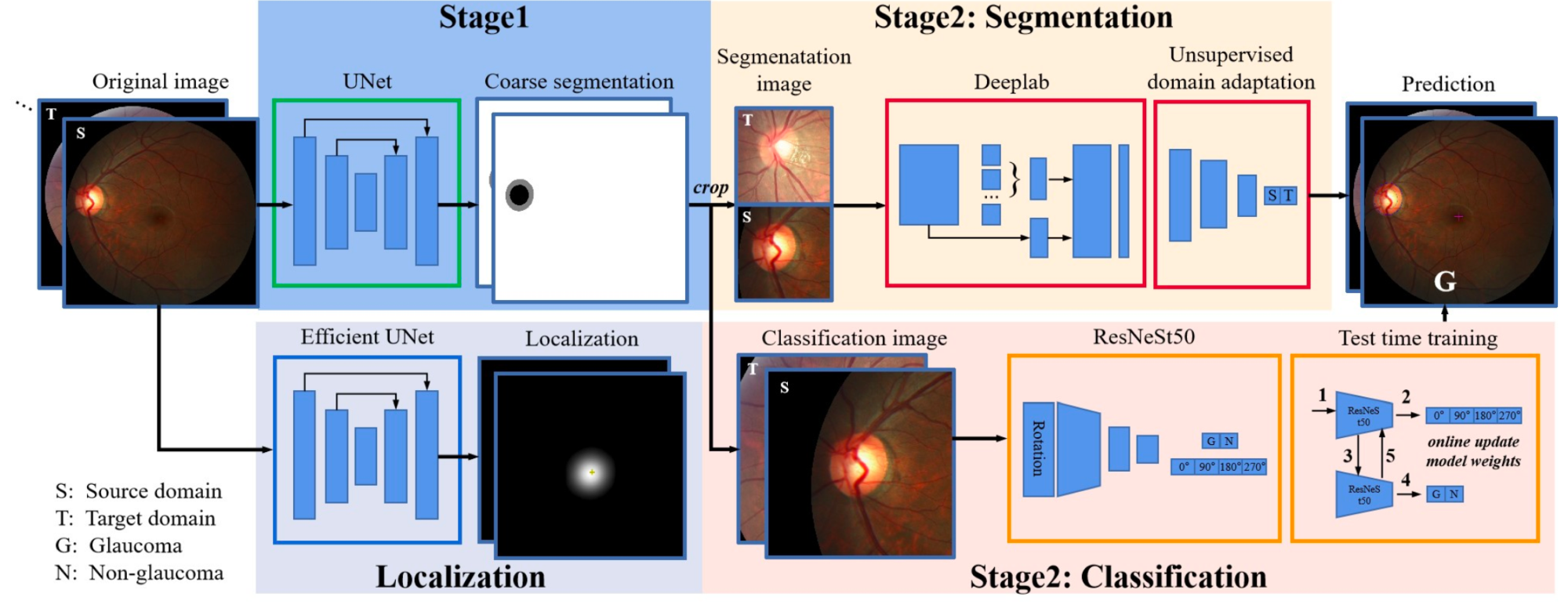}
\caption{An overview of the proposed framework of the MAI team.}
\label{fig:MAI-task1}
\end{figure}

\begin{figure}[tb]
\centering
\includegraphics[width=\linewidth]{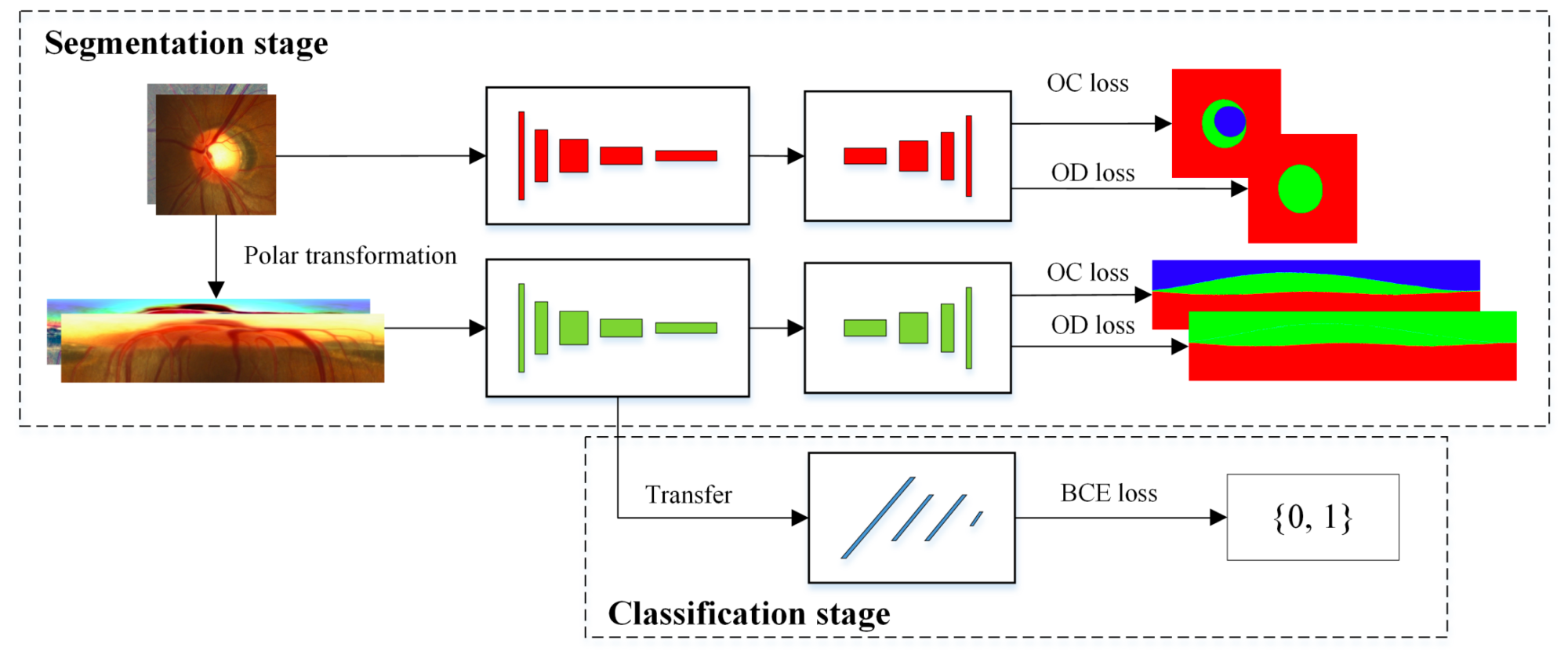}
\caption{Segmentation and Classification stages procedure of the Pami-G team.}
\label{fig:Pami-G-task2}
\end{figure}

The MIAG ULL team first localized the OD region and cropped it (\cite{sigut2017contrast}), then VGG19 was used to predict the glaucoma probability. The ALISR team extracted the OD patches by using Mask R-CNN (\cite{he2017mask}) and classified the patches by using cross-stage partial network (CSPNet) (\cite{wang2020cspnet}). The Pami-G team transferred the knowledge of the FCN encoder trained in the OD/OC segmentation task. As shown in Fig.~\ref{fig:Pami-G-task2}, they removed the layers of up-sampling module of the FCN, and connected three fully connected (FC) layers after the encoder layers of the FCN. During training, only the FC layers parameters were learned.

The CBMIBrand team first detected the center of OD by using Mask R-CNN (\cite{he2017mask}), and then cropped multiple regions with various scales enclosing the OD ($384 \times 384, 416 \times 416, 448 \times 448, 480 \times 480,512 \times 512$ pixels) to solve the problem of inconsistent object size caused by the inconsistent image size in the training and testing sets. As for the classification, they devised an ensemble strategy to train six models, including ResNet18 (\cite{7780459}), ResNer34, ResNet50, ResNet101, DenseNet121 (\cite{huang2017densely}), and DenseNet169,  separately and finally average their outputs.

The remaining team MIG utilized two ResNet50 frameworks to predict the glaucoma probabilities based on both the whole fundus images and the cropped OD patches. Besides, considering that the model was easy to overfit and had poor generalization, they modified the ResNet50 model by replacing the fully connected layer with global average pooling (GAP) to improve the robustness. And, they also designed a $1 \times 1$ convolutional compression feature channel (CC) to reduce the overfitting. They adopted the designed ResNet50\_2GAP, ResNet50\_3GAP (where 2GAP and 3GAP represent the fusion of the last two or three scales of the model, respectively), and ResNet50\_CC to predict the probabilities of the glaucoma based on the cropped OD patches. Finally, the output of the ensemble of the above five models was averaged.

\subsection{Segmentation of optic disc and cup}
The frameworks proposed by the teams can be divided in two categories: segmenting OD/OC directly, and segmenting OD/OC from coarse to fine. The ALISR team designed a variant of DeepLab-v3 (\cite{chen2017deeplab}), which used a retrained EfficientNet-B1 (\cite{tan2019efficientnet}) to replace the ResNet architecture of the deep convolutional neural network (DCNN) backbone in the original DeepLab-v3. Then the network was used to segment the OD/OC directly. The VUNO EYE TEAM used a two-branch network to segment the OD and OC using fundus image and vessel image (\cite{fang2022adam}). Since OC and OD occupy a relatively small proportion of fundus images, and the OC is inside the OD, many teams first detected the coarse area of OD, and then segmented the fine OD and OC in this area. The CBMIBrand team first used the detection branch of Mask R-CNN to detect OD regions, and then used the segmentation branch of Mask R-CNN to segment OD and OC regions simultaneously in the feature map of OD. The teams cheeron, TeamTiger, MIG and MAI all segmented the OD first using U-Net, and then the cheeron team adopted a ResU-Net to further segment the OD and OC, while the TeamTiger adopted a generative adversarial network. The MIG team adopted a CE-Net (\cite{gu2019net}), which is based on the U-Net model with ResNet34 as encoder, to segment the fine OD/OC. The MAI team utilized Deeplab-v3+ framework to achieve the precise segmentation of OC/OD. Moreover, they adopted a classical unsupervised domain adaptation strategy (\cite{tsai2018learning}) to maintain the segmentation performance on different devices, which is introduced in Section \textit{3.2 Strategies for domain adaptation} in the paper. In addition to the general binary cross-entropy loss, the MAI team also designed an intersection over union (IOU) based loss term, i.e., $loss_{additional} = 1 - IOU$.

The MIAG ULL and the Pami-G teams both considered the anatomical relationship of the OD and the fovea. The MIAG ULL team utilized PSPNet (\cite{zhao2017pyramid}) with ResNet50 as encoder to segment the macular region and the coarse OD region. Subsequently, two PSPNet models with ResNet50 as encoder were used to achieve the fine segmentation of OD and OC. The Pami-G team designed two models to learn the segmentation and localization of the OD and macular regions simultaneously, which will be described in Section~\ref{fovea_loc}. Then they extracted a square ROI of OD with the length setting to $0.8 \times L$, where $L$ was the distance between the center of the OD and the macular region. As shown in the segmentation stage of Fig.~\ref{fig:Pami-G-task2}, they introduced polar transformation to transform the ROI patches, and then put the original ROI as well as the polar transformed ROI patches into two FCNs which had the same structure. It is worth noting that the Pami-G team also preprocessed the image, i.e., the background brightness of the fundus image was estimated by Gaussian filter, which was then subtracted to balance the luminance and enhanced the contrast of the whole image. They designed two loss terms according to the cases that only segment OD region and segment OD and OC regions simultaneously. Specifically, $p_0, p_1, p_2$ represented the predicted background map, the predicted OD mask excluding OC mask, and the predicted OC mask, respectively, and $g_0, g_1, g_2$ represented the corresponding ground truth. In this way, $p^{'}_{0}=p_{0}, p^{'}_{1}=p_{0}+p_{1}$ represented the predicted background map and the OD mask, and $g^{'}_{0}=g_{0}, g^{'}_{1}=g_{0}+g_{1}$ was the corresponding ground truth. They set 
\begin{equation}
\small
    Loss_{OC}=(C-\sum_{c=0}^{2}\frac{\sum p_{c}g_{c}}{\sum p_{c}g_{c}+\alpha \sum (1-p_{c})g_{c}+\beta \sum p_{c}(1-g_{c})}),
\end{equation}
\begin{equation}
\small
    Loss_{OD}=(C^{'}-\sum_{c=0}^{1}\frac{\sum p^{'}_{c}g^{'}_{c}}{\sum p^{'}_{c}g^{'}_{c}+\alpha \sum (1-p^{'}_{c})g^{'}_{c}+\beta \sum p^{'}_{c}(1-g^{'}_{c})}),
\end{equation}
where $\alpha$ and $\beta$ were the trade-offs of penalties and all were set to 0.5 in their experiments. The total loss in their framework was as follows:
\begin{equation}
    Loss_{T}=Loss_{OD} + fLoss_{OC},
\end{equation}
where $f=1$ if and only if OC annotation existed, otherwise $f=0$. As a result, during inference stage, they utilized polar transformed patch to segment OD and Cartesian coordinate patch to segment OC.

\begin{figure}[tb]
\centering
\includegraphics[width=\linewidth]{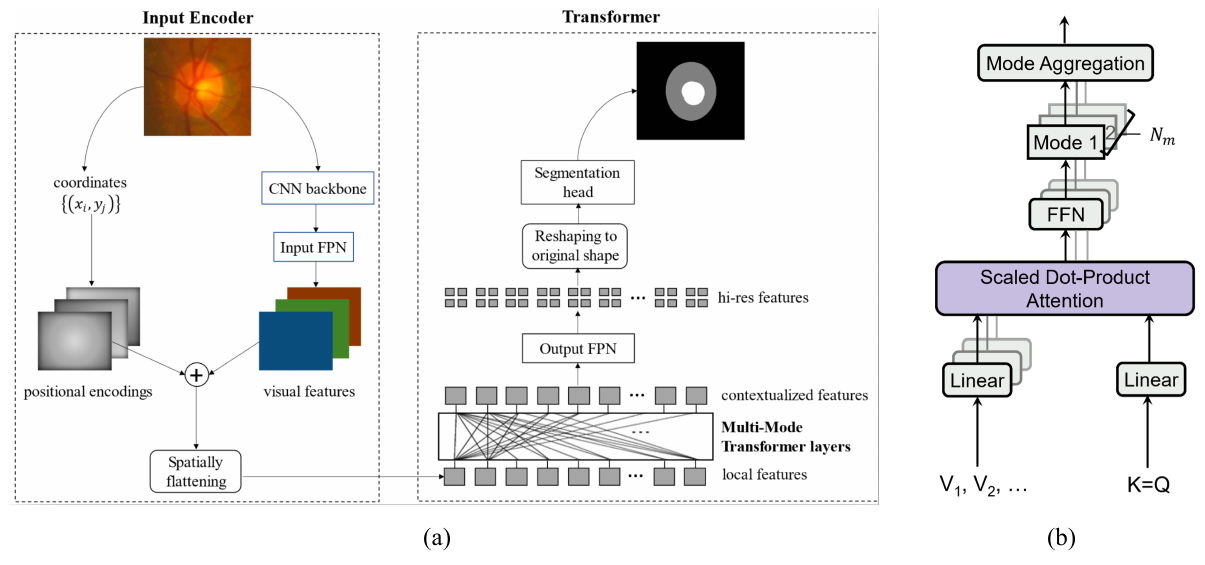}
\caption{(a) The framework of the EyeStar Team in task 2. (b) Multi-Mode Transformer layers. $Q$ and $K$ were tied in each mode, and each mode had its private attention, key projection matrix and FFN.
}
\label{fig:EyeStar-task2}
\end{figure}

The EyeStar team adopted a vision transformer method, and to increase efficiency, the transformer took a coarse feature maps from a CNN backbone as input. As shown in Fig.~\ref{fig:EyeStar-task2}(a), the output feature maps of the transformer were upsampled with a feature pyramid network (FPN) before being classified by a segmentation head. In their experiments, EfficientNet-B4 (\cite{tan2019efficientnet}) was used as the CNN backbone; accordingly, the number of feature channels was set as 1792. To increase the spatial resolution of the feature maps, they adopted an input FPN and an output FPN that upsample the feature maps at the transformer input and output, respectively. The positional encoding was a learnable sinusoidal. Given a pixel coordinate $(x, y)$, the $C$-dimension positional encoding vector $p(x, y)$ is:
$$p_{i}(x, y)=\left\{
\begin{aligned}
sin(a_{i}x + b_{i}y + c_{i}), & if & i < C/2 \\
cos(a_{i}x + b_{i}y + c_{i}), &if & i \geq C/2
\end{aligned}
\right.
$$
where $C$ was the same as the channel of the CNN backbone, $i$ indexed the elements in $p$, $\{a_{i}, b_{i}, c_{i}\}$ were learnable weights. The $(x, y)$ coordinates were normalized to [0,1] to maintain a consistent behavior across different image sizes. The visual features and positional encoding of the whole image were summed up before being fed to the transformer: $X_{vol}=f(X_{0})+p(X_{0})$, where each image unit (a small downsampled patch) corresponded to a $C$-dimensional vector. 
The core of the transformer layer was self-attention. To improve self-attention for image applications, they proposed a novel Multi-Mode Transformer Layer. Specifically, it had one attention head and $N_{m}$ value transformations, outputting $N_{m}$ groups of features:
\begin{equation}
\label{Q}
    Q = XW_{Q},
\end{equation}
\begin{equation}
\label{eq:att}
    Att\_weight(X, X) = softmax(\frac{QQ^T}{\sqrt{d_{k}}}),
\end{equation}
\begin{equation}
    V^{(k)} = XW^{(k)}_{V},
\end{equation}
\begin{equation}
    Mode^{(k)} = Att\_weight(X, X)\centerdot V^{(k)},
\end{equation}
\begin{equation}
    X^{(k)}_{out} = FFN^{k}(Mode^{(k)}), with\quad k\in {1,...,N_{m}},
\end{equation}
where all modes shared the attention weight matrix as computed by Eq.~\ref{eq:att}, but each mode had a private value matrix $W^{(k)}_{V}$ and FFN. FFN is a two-layer feed-forward network with residual connections to further transform the fused features. As shown in Fig.~\ref{fig:EyeStar-task2}(b), the features from all modes were aggregated with dynamically computed mode attention $G$, which was inspired by the Split Attention~\citep{zhang2020resnest}:
\begin{equation}
    B^{(k)}=X^{(k)}_{out}W_{G} + C_{G},
\end{equation}
\begin{equation}
\label{eq:B}
    G = softmax(B^{(1)},...,B^{(N_m)}),
\end{equation}
\begin{equation}
\label{eq:X_out}
    X_{out} = G\centerdot (X^{1}_{out},...,X^{(N_{m})}_{out})^T ,
\end{equation}
where $W_{G}$, $C_{G}$ were the parameters of the linear layer. In Eq.~\ref{eq:B}, the softmax probabilities were normalized over the $N_{m}$ modes, and Eq.~\ref{eq:X_out} took a weighted sum over the mode features. At last, the segmentation head was simply a $1\times1$ convolutional layer, whose number of output channels was 3 for the OD/OC segmentation task.

\subsection{Fovea localization}
\label{fovea_loc}
The fovea localization task is to predict the coordinate $(x, y)$ of fovea, which is the center of the macular region. 
We can divide the methods into the following three categories: direct using the original image, using the relative position strategy of the OD and the fovea, and using the coarse-to-fine strategy. 

The VUNO team directly predicted a single pixel of fovea segmentation mask, and two deviation masks on the x- and y-axis, the same as they used in the ADAM Challenge (\cite{fang2022adam}). The MAI team and ALISR team transformed the fovea localization task into a distance map regression, in which the MAI team utilized a U-Net with EfficientNet-B5 as the feature extractor, and the ALISR team utilized a pre-trained U-Net (\cite{meyer2018pixel}) to achieve the regression. The cheeron team transformed the fovea localization into an object detection task, and used the latest YOLOv5 to predict its bounding box.

\begin{figure}[tb]
\centering
\includegraphics[width=\linewidth]{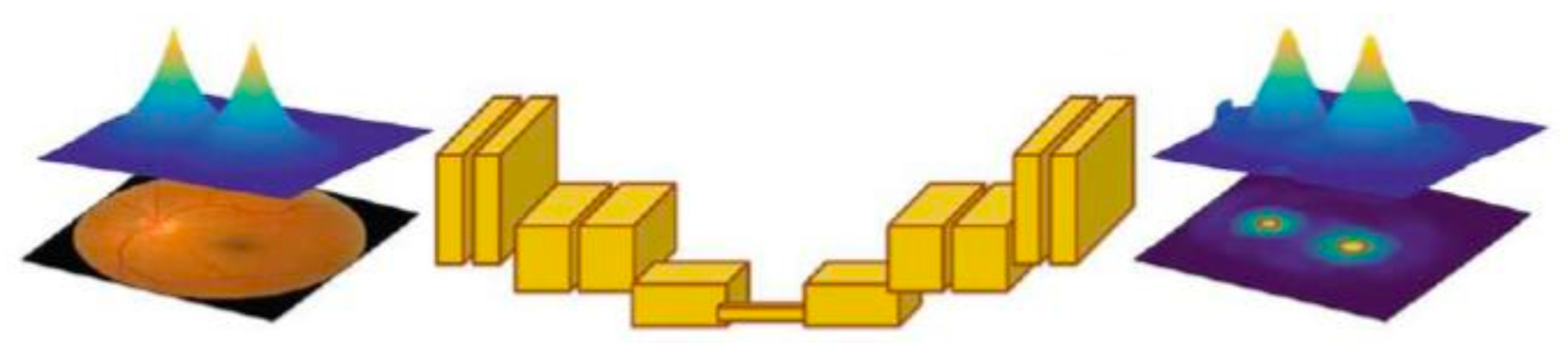}
\caption{The method of the MIG team in task 3 for joint fovea and OD localization via regressing a Bi-Distance map.}
\label{fig:MIG-task3}
\end{figure}

\begin{figure}[tb]
\centering
\includegraphics[width=\linewidth]{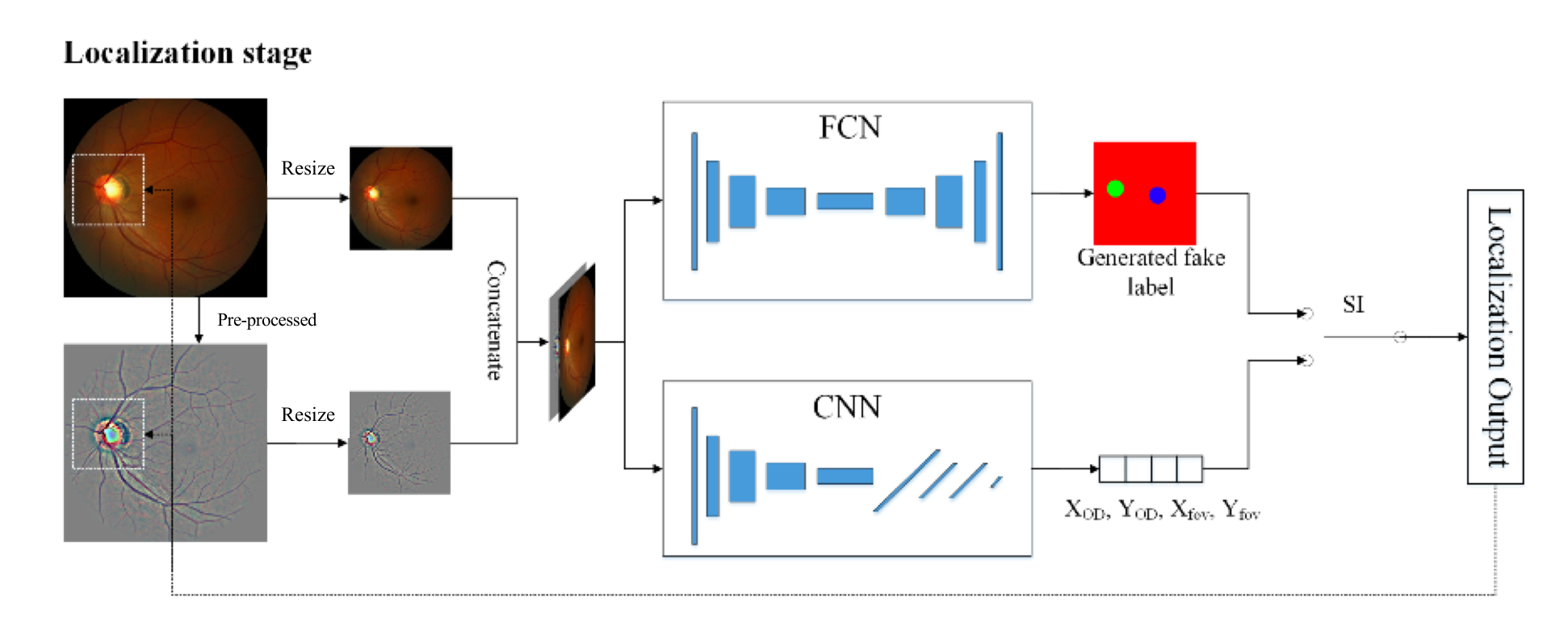}
\caption{The architecture of the Pami-G team in Task3.}
\label{fig:Pami-G-task3}
\end{figure}

The MIG team believed that joint learning of the position of each pixel associated with OD and fovea will help to automatically understand the overall anatomical distribution (\cite{meyer2018pixel}), and they also transformed the localization task into a distance map regression problem. As shown in Fig.~\ref{fig:MIG-task3}, they designed a Bi-Distance map for each pixel $(x, y)$, and adopted a U-Net to predict the map. In the Bi-Distance map, pixel value $B(x, y)$ is the distance from the nearest landmark of interest (the center of OD or the fovea). The MIAG ULL also considered the relationship between the OD and the fovea, they adopted ResNet50 and PSPNet to segment the OD and fovea simultaneously. Similarly, the Pami-G team also segmented these two fundus structures, but they utilized an FCN framework (see Fig.~\ref{fig:Pami-G-task3}). In addition, they also used another CNN branch to predict the center of the OD and the fovea. They designed a shape index (SI) to determine which fovea prediction was the final result. SI was defined as $SI=C^2/S$, where $S$ and $C$ denote the area and the perimeter of a region in prediction. When $SI \in (11, 12.2)$, they used the center of FCN output region as the localization result, and they used the CNN output when $SI \notin (11, 12.2)$.

\begin{figure}[tb]
\centering
\includegraphics[width=\linewidth]{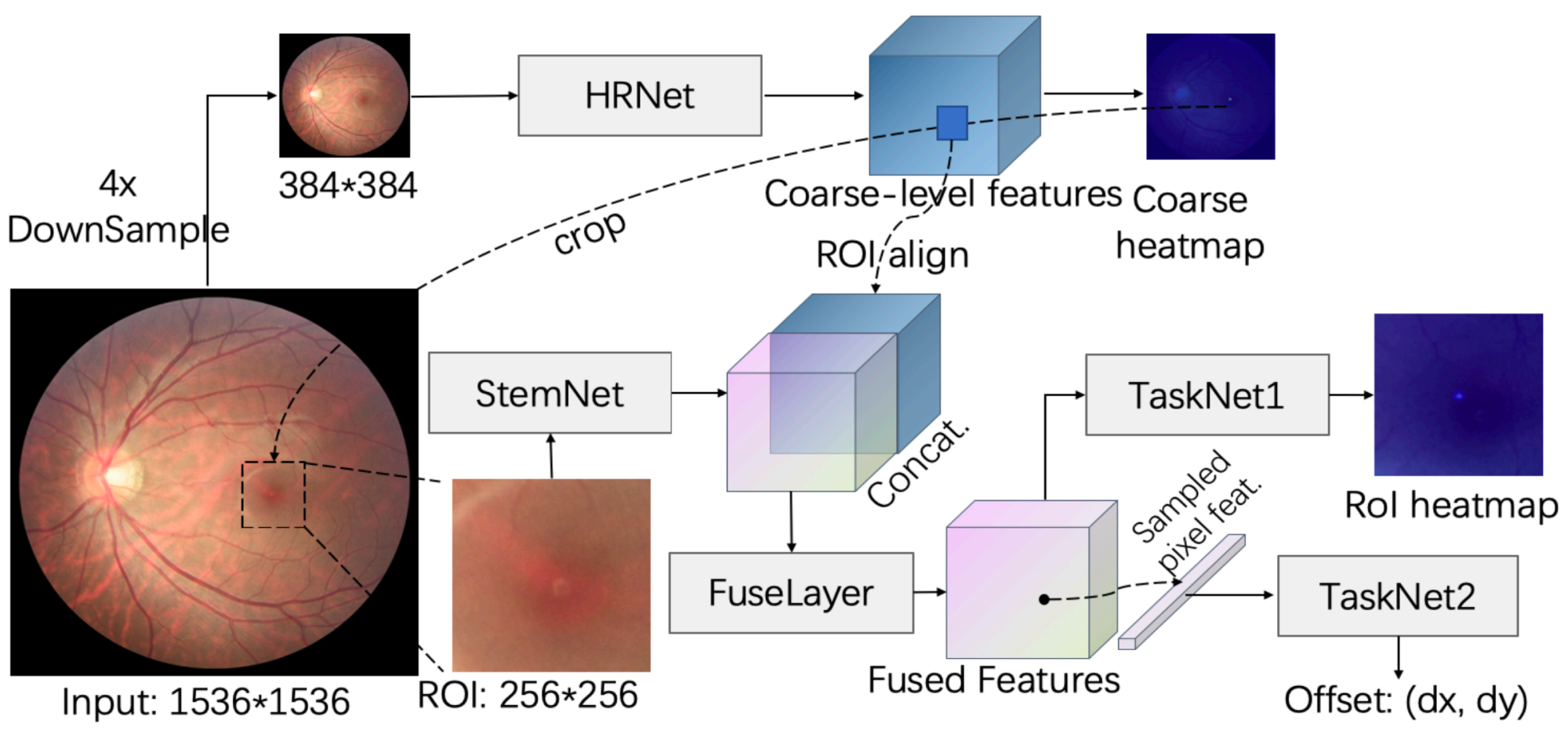}
\caption{The architecture of the EyeStar team in Task3.}
\label{fig:EyeStar-task3}
\end{figure}

The TeamTiger designed a coarse-to-fine method. The initial model was an EfficientNet-based to get a tentative position of the fovea, and the second model worked on the patch around the identified coarse fovea area to segment the macular region. The second model was U-Net with EfficientNet as encoder. The center of the fovea segmentation mask is the fovea localization result. Similarly, the CBMIBrand team also adopted the coarse-to-fine strategy. They first cropped the coarse fovea patches according to the positional relationship between the OD center and the fovea center. Specifically, they obtained the center of OD based on task 2. Then, they approximated the fovea position by searching the region directing down of 1/6 OD diameter (ODD) and right/left of 2.5 ODD (right and left eyes) starting from the OD center. Finally, they cropped multi-scale fovea ROIs ($384 \times 384, 448 \times 448, 480 \times 480,512 \times 512, 640 \times 640$ pixels), and resized them into $512 \times 512$ pixels. They concatenated the multi-scale patches at the channel dimension and finally adjusted DenseNet to perform a fovea coordinate regression. Analogously, the EyeStar team adopted the coarse-to-fine strategy, but they fused the local feature of the whole image and the global feature of the cropped patch. As can be seen in Fig.~\ref{fig:EyeStar-task3}, the input image was down-sampled by $4 \times$ and fed into a pre-trained HRNet (\cite{sun2019deep}) to get the coarse predicted heatmaps. A ROI region on the original input image, centered at the peak pixel of the predicted heatmaps, was then cropped and fed into a StemNet (\cite{sun2019deep}) to obtain the fine-scale features. The fine-scale features were concatenated with the pixel-aligned coarse-level features (using the ROIAlign layer (\cite{he2017mask})), and processed by FuseLayer. The fused features are finally passed through the TaskNet1 to get fine-scale heatmap and through the TaskNet2 to predict the offset of the sampling location to the ground truth. In their framework, the FuseLayer was a convolutional block that outputs 32 feature channels. The TaskNet1 consisted of two convolutional blocks with 32 and 1 channels. TaskNet2 was a multi-layer perceptron with 32, 16 and 2 channels.

\section{Supplementary figures}

\setcounter{figure}{0}  
\begin{figure}[!tbh]
\centering
\includegraphics[width=\linewidth]{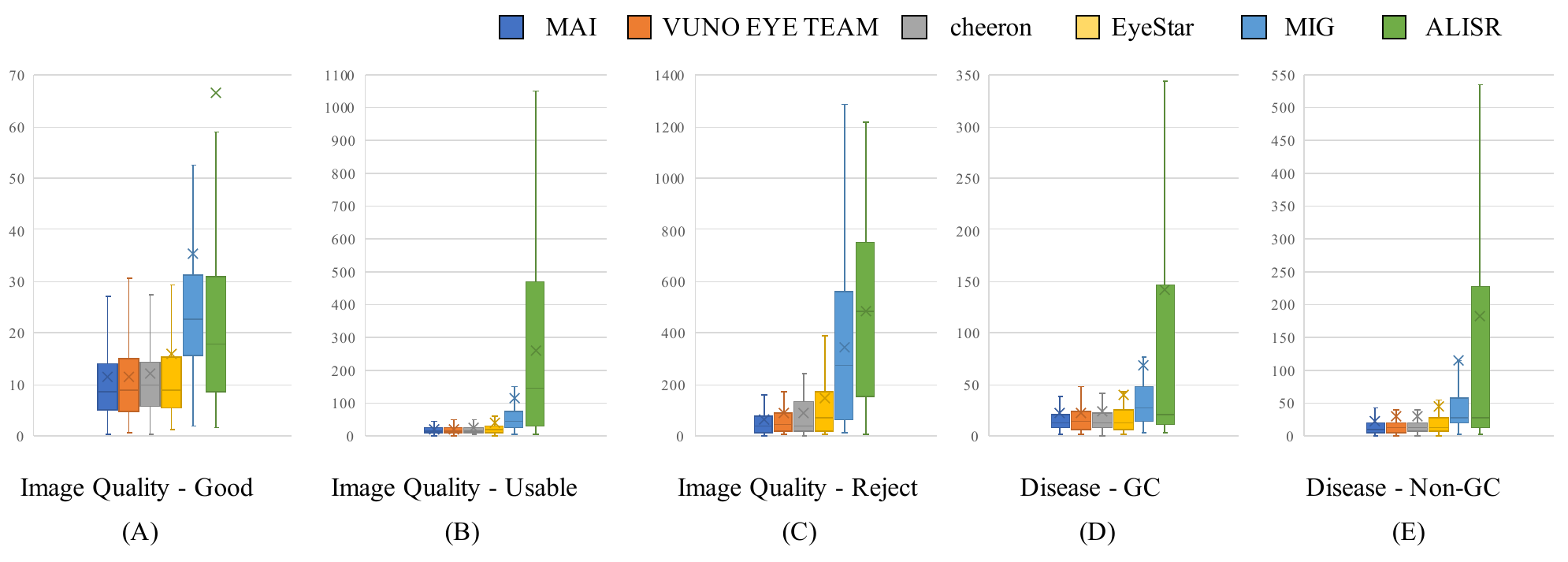}
\caption{The complete boxplots corresponding to Fig. 10 in the paper, represents the fovea localization results stratified by data image quality and disease situations. (A)-(C) correspond to samples with ‘good', ‘usable', and ‘reject' image quality, and (D)-(E) correspond to glaucoma samples and non-glaucoma samples.}
\label{fig:diff-task3-supplemental}
\end{figure}

\begin{figure}[!tbh]
\centering
\includegraphics[width=0.8\linewidth]{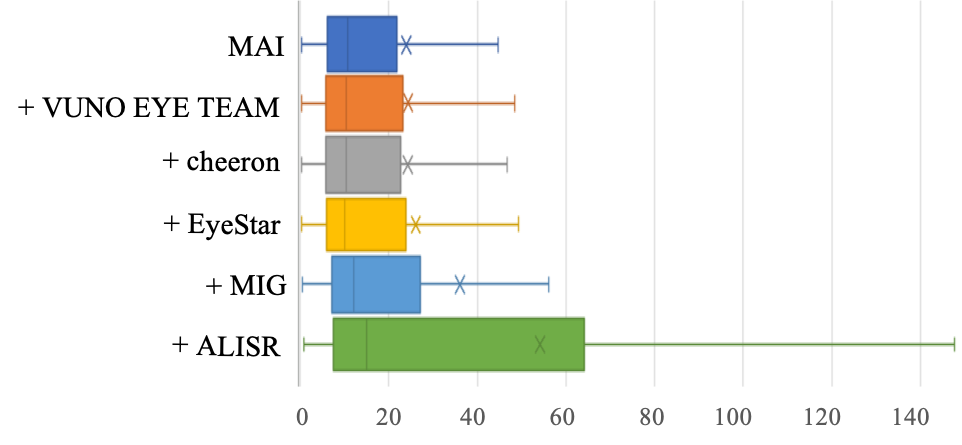}
\caption{The complete boxplot corresponding to Fig. 11(C) in the paper, represents the evaluation of the ensemble results in the foveal localization task.}
\label{fig:ensemble-task3-supplemental}
\end{figure}

\bibliographystyle{unsrtnat}
\bibliography{references}

\end{document}